\documentclass[%
unsortedaddress,
preprint,
 amsmath,amssymb,
 aps,
prb,
]{revtex4-1}

\usepackage{graphicx}
\usepackage{dcolumn}
\usepackage{bm}
\usepackage{multirow}

\usepackage{xcolor}

\begin{document}

\title{Diffusion Monte Carlo Study of the O$_2$ Adsorption on a Single Layer Graphene}
\author{Hyeondeok Shin}%
\affiliation{Computational Science Division, Argonne National Laboratory, Argonne, Illinois 60439, USA}
\author{Ye Luo}%
\affiliation{Computational Science Division, Argonne National Laboratory, Argonne, Illinois 60439, USA}
\author{Anouar Benali}
\email{benali@anl.gov}
\affiliation{Computational Science Division, Argonne National Laboratory, Argonne, Illinois 60439, USA}
\author{Yongkyung Kwon}
\email{ykwon@konkuk.ac.kr}
\affiliation{Department of Physics, Konkuk University, Seoul 05029, Korea}


\date{\today}

\begin{abstract}
Diffusion Monte Carlo (DMC) calculations were performed for an accurate description of the nature of the O$_2$ adsorption on a single layer graphene. We investigated the stable orientation of O$_2$ at a specific adsorption site as well as its equilibrium adsorption energy. At equilibrium adsorption distances, an O$_2$ molecule was found to prefer a horizontal orientation, where the O-O bond is parallel to the graphene surface, to the vertical orientation. However, the vertical orientation is favored at the O$_2$-graphene distances shorter than the equilibrium distance, which could be understood by the steric repulsion between O and C atoms. Contrary to previous DFT calculations, our DMC calculations show that the midpoint of a C-C bond (a bridge site) is energetically preferred for the O$_2$ adsorption to a center of a hexagonal ring (a hollow site). 
The lowest DMC adsorption energy was found at an intermediate point between a hollow and a bridge site, where the O$_2$ adsorption energy was estimated to be $-0.142(4)$ eV that was in very good agreement with the recently-reported experimental value.
Finally, we have found that O$_2$ is very diffusive on the surface of graphene with the diffusion barrier along a bridge-hollow-bridge path being as small as $\sim 11$ meV.
\end{abstract}

\pacs{Valid PACS appear here}

\maketitle


\section{\label{sec:level1}Introduction}
Graphene has drawn much interest from various areas of chemistry, physics and industry because of its unique electronic and optical properties.\cite{novoselov05,zhang05,geim07,geim09}
Among others, graphene-based materials have been identified as promising candidates for future enhanced gas sensor, beyond carbon nanotubes\cite{kong00} or nanowires,\cite{cui01} because of their high reactivity to adsorbates.
An adsorbed gas molecule on a graphene surface was shown to play a fundamental role as a donor/acceptor and to alter significantly its transport properties.\cite{schedin07,lu09,varghese15} 
Additionally, its atomically-thin two-dimensional hexagonal structure allows all carbon atoms to work as surface atoms for molecular adsorbates, yielding a large sensing area per volume.
A number of theoretical and experimental studies have recently attempted to quantify and describe the performance of a graphene-based material, including a pristine graphene sheet, as a gas detector.\cite{wehling08,lee13,chen14,ganji15} 
For example, an exfoliated graphene on SiO$_2$ was reported to be capable of detecting NO$_2$, a well-known critical air pollutant, at very low concentrations of a few parts per billion.\cite{schedin07}
Other graphene-based materials, such as carbon nanotubes, graphene oxides and doped graphene, have also been considered as sensing substrates to detect not only NO or NO$_2$ but other gases such as NH$_3$, CO$_2$ and O$_2$.\cite{ratinac10,yuan13,jaaniso14,novikov16}
The adsorption energy of a molecule on a substrate is a key metric in identifying a gas-sensing material.  
In the past decades, first-principle  density functional theory (DFT) calculations have been widely used to investigate energetics of different gas adsorptions on the surface of a graphene-based material.\cite{huang08,leenaerts09,zhou11,choudhuri15,liang17} 
However, the DFT results were often strongly dependent on the choice of exchange-correlation functional, which can be attributed to some limitation of its Kohn-Sham mean-field scheme in describing subtle competition between the covalent nature of the adsorbate-substrate interaction and its van der Waals (vdW) nature.

Oxygen dimer (O$_2$) is the second most abundant molecule in the atmosphere. Among other usages, O$_2$ can be used to control the rate of combustion or to destruct hazardous and waste materials in incinerators. It needs to be removed from the atmosphere to avoid corrosion. Because of the crucial importance in its industrial application, the adsorption nature of O$_2$ has been intensively studied for various sensing materials including carbon allotropes.\cite{kauffman09,rajavel15,tang17}
Recently, the adsorption energy of O$_2$ on graphene was experimentally measured using temperature-programmed terahertz emission microscopy.\cite{begsican17} However, detailed information about the adsorption site and equilibrium distance along with its preferred molecular orientation is yet to be accurately determined.

On the theoretical side, several DFT calculations were performed to study the adsorption of O$_2$ on graphene. 
While a geometry relaxation of Yan {\it et~al.}~\cite{yan12} based on a GGA functional showed the adsorption of O$_2$ at a hollow site, the center of a hexagonal ring on the graphene surface, with the adsorption energy of -0.13 eV, Guang {\it et~al.} used a GGA functional with the addition of empirical vdW dispersion of Grimme to find an O$_2$ adsorption at a bridge site, the midpoint of a C-C bond,
with the adsorption energy of -0.15 eV~\cite{guang13}.
On the other hand, two independent LDA calculations predicted the adsorption of O$_2$ at a hollow site~\cite{zou11,kaur18}. This discrepancy in previous DFT calculations asks for
a more systematic theoretical study for the O$_2$ adsorption on graphene with a  computational method that fully incorporates electron-electron correlations.

Quantum Monte Carlo (QMC) can provide accurate estimations of the adsorption energy by solving a many-body Schr\"{o}dinger equation stochastically. QMC was used successfully to study various vdW-dominated systems and provided accurate description of vdW interactions in molecular systems, bulk solids\cite{benali14}, and 2D materials including bilayer graphene and other low-dimensional carbon allotropes.\cite{shin14,mostaani15,shin2017}   
In this study, we report diffusion Monte Carlo (DMC) results for the energetics, the adsorption distances, and preferred orientations of an O$_2$ molecule adsorbed at different sites of a single layer graphene. We then estimate the O$_2$ diffusion barrier on the graphene surface using both DMC and vdW-corrected DFT calculations. 

\section{\label{sec:level2}Computational details}
Our QMC calculations were carried out with the  fixed-node DMC method\cite{Reynolds1982,foulkes01} as implemented in the QMCPACK code.\cite{QMCPACK}
Slater-Jastrow trial wave functions were used in the DMC algorithm with up to three-body Jastrow functions in order to include ion-electron (one-body), electron-electron (two-body), and ion-electron-electron (three-body) correlations.
All DMC calculations used a time step of 0.005 Ha$^{-1}$ and size-consistent T-moves for the variational evaluation of the non-local pseudopotentials.\cite{casula10}
In order to reduce finite-size effects, 
DMC total energies of four different supercells averaged over 36 twist angles (twist-averaged boundary conditions\cite{lin01}) were extrapolated to the bulk limit. 
For atomic calculations of both carbon and oxygen, we used norm-conserving scalar-relativistic energy-consistent pseudopotentials developed by Burkatzki, Filippi, and Dolg\cite{burkatzki07,burkatzki08}, whose accuracies were confirmed in various  DMC studies on similar materials.\cite{shin14,benali16,luo16,shin2017,shin17}.
The single Slater determinant in the wave function was constructed with single-particle orbitals obtained through spin-polarized DFT calculations based on the Perdew-Burke-Ernzerhof (PBE) parameterization\cite{perdew96}, which were done with the QUANTUM ESPRESSO code\cite{QE}.
To analyze effects of the dispersion force on the O$_2$ adsorption, DFT calculations were also performed using several vdW-corrected functionals; Grimme (DFT-D2) correction to PBE exchange correlation (XC) functional\cite{grimme04,grimme06,grimme10}, self-consistent vdW-corrected XC functionals and more recent rVV10 correction.\cite{dion04,lee10,sabatini13} 
In order to minimize the interactions between O$_2$ and its periodic images, we used a $3 \times 3 \times 1$ single layer graphene with one adsorbed O$_2$ molecule in its triplet state. The distance between O$_2$ and its first periodic image is around 7~$\text{\AA}$, which corresponds to the dissociation limit of O$_2$-O$_2$ dimer.\cite{aquilanti99,lamoneda05} 
Self-consistent spin-polarized DFT calculations were done using 250 Ry plane-wave basis-set cut-off and $6 \times 6 \times 1$ Monkhorst-Pack $k$-point grid mesh.\cite{monkhorst76}
The O-O bond length in O$_2$ was chosen to be 1.208 \text{\AA} from the experimental result for the equilibrium O-O bond length\cite{lide03} and the experimental value of 1.421 \text{\AA} for the C-C bond length of graphite was used for that of a single layer graphene.\cite{lynch66}

\section{\label{sec:results}Results}

\begin{figure}[t]
 \includegraphics[width=6 in]{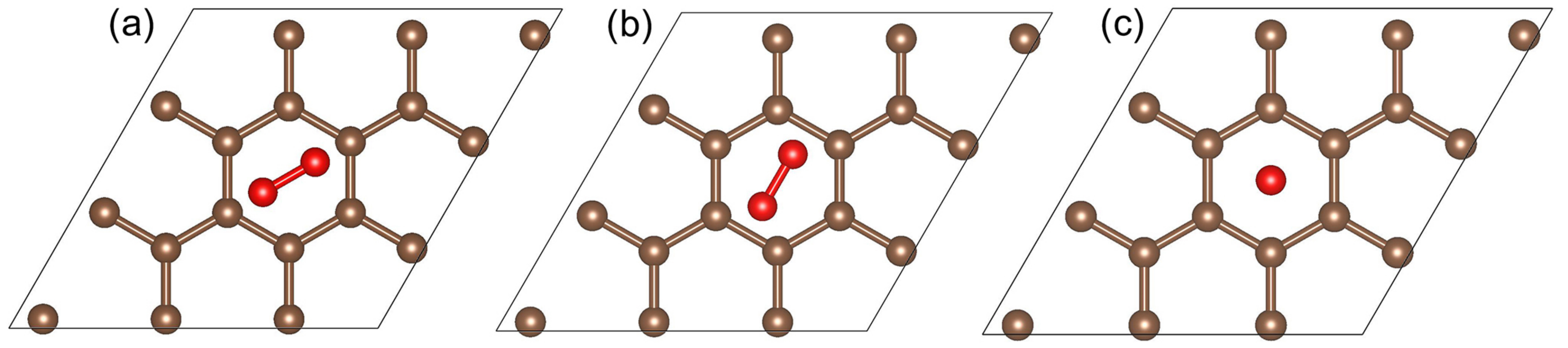}
 \caption{Top view of an O$_2$ molecule in three different orientations adsorbed at a hollow site on the graphene surface; (a) O$_2$ parallel to the graphene surface with O atoms pointing at the carbon atoms ($A$ mode), (b) O$_2$ parallel to the graphene surface with O atoms pointing at the middle of a C-C bond ($B$ mode), and (c) O$_2$ perpendicular to the graphene surface ($V$ mode).}
 \label{fig:O2_mode}
\end{figure}

We first investigate the adsorption of a single O$_2$ molecule at a hollow site, which was reported in previous DFT studies~\cite{zou11,yan12,kaur18} to be a stable adsorption site for O$_2$.
To determine the preferred orientation of O$_2$ adsorbed at a hollow site, 
we here consider its three different orientation modes as shown in Figure~\ref{fig:O2_mode}; the O$_2$ molecular axis is oriented in the directions parallel to the graphene surface in the first two modes called $A$ and $B$ modes but it is perpendicular to the graphene surface in the $V$ mode.
Ideally, the most stable mode could be determined through a geometry optimization of the O$_2$-graphene complex. 
A robust DMC algorithm to compute forces, however, has yet to be available.  
Therefore, we reduce the degrees of freedom of the system only to a vertical translation of O$_2$ with respect to the graphene surface and then compute DMC energies as a function of the distance from the surface. This allows us to construct a binding curve for each orientation mode. 
Because a localized puckering of the graphene surface is expected to occur near the adsorption site of O$_2$, we performed a full DFT optimization of the adsorbant and the graphene surface using rVV10 vdW-corrected exchange-correlation functional to assess the level of deformation and its associated energy.
It was found that geometries and total energies varied only a little, less than 0.02~\AA~and 10 meV, respectively, with the structural deformation. 
From this we conclude that our approach of fixing the geometry of the graphene sheet and the O-O bond length of O$_2$ introduces a rather small bias that should not affect the quality of the calculations or the conclusions of this study.

\begin{figure}[t]
 \includegraphics[width=6 in]{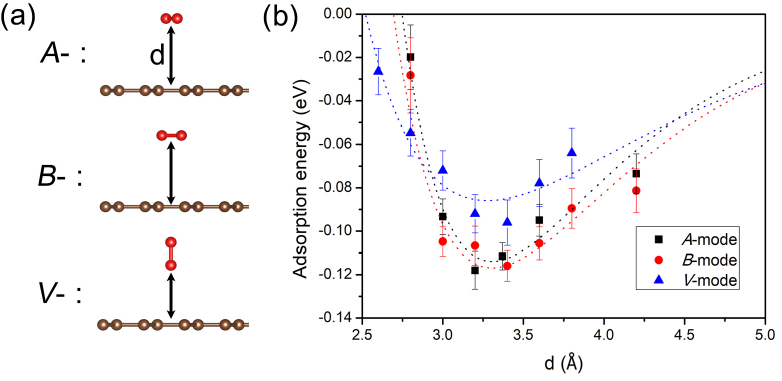}
 \caption{(a) Side view of $A$, $B$, and $V$ orientation mode of O$_2$ adsorbed at a hollow site in a graphene sheet. Red and brown sphere indicate oxygen and carbon atoms, respectively. (b) DMC adsorption energy of O$_2$ as functions of the distance from the graphene surface for the three orientation modes. The dotted lines represent the Morse function fits.}
 \label{fig:DMC_tot}
\end{figure}

Following the scheme described above,
we performed DMC calculations to compute the adsorption energy of a single O$_2$ molecule, defined by
\begin{equation}
E({\rm graphene},{\rm O}_{2}) - E({\rm graphene}) - E({\rm O}_{2}) 
\end{equation}
where $E({\rm graphene},{\rm O}_{2})$, $E({\rm graphene})$, and $E({\rm O}_{2})$ are, respectively, 
the total energies of graphene with the O$_2$ adsorbate, the pristine graphene, 
and an isolated O$_2$ molecule. 
Figure~\ref{fig:DMC_tot} shows DMC adsorption curves of O$_2$ at a hollow site that were simulated with a $2 \times 2 \times 1$ supercell involving 4 O$_2$ molecules and 72 C atoms. 
As can be seen, the equilibrium binding distances are $3.2$ to $3.4$~\AA~and are quite similar among the three orientation modes.
While equilibrium binding energies of $A$ and $B$ modes are indistinguishable from each other within error bars, the $V$ mode turns out to be the least probable one among the three orientations.
By choosing the mode with the lowest adsorption energy at a specific distance from the binding energy curves of Fig.~\ref{fig:DMC_tot}(b), we determine the preferred orientation profiles of an O$_2$ molecule at a hollow site; O$_2$ is found to prefer an orientation parallel to the graphene surface at the distances longer than 3.0~\AA~from the graphene surface while the vertical $V$ mode is favored at shorter distances. 
It is not surprising to find that the $V$ mode is most stable at short distances
because O$_2$ in this mode will be less affected by the steric repulsion between C and O atoms when approaching the graphene surface. This mode is expected to be the preferred orientation when O$_2$ crosses through the graphene layer. 

\begin{figure}[t]
 \includegraphics[width=6 in]{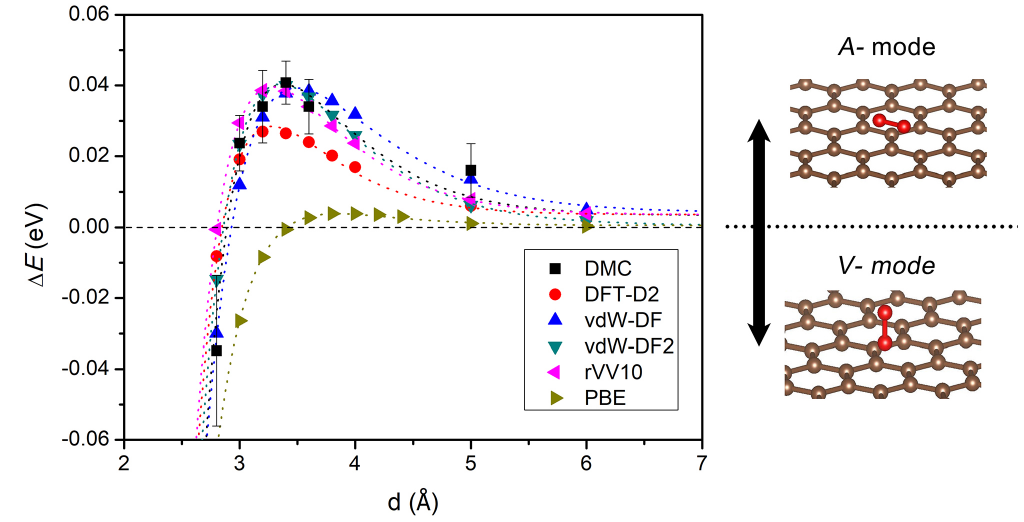}
 \caption{Total energy difference between the $A$ and the $V$ modes as functions of the distance between the O$_{2}$ molecule and the graphene surface, {\it i.e.}, $\Delta E = E_{V}(d)-E_{A}(d)$ where $E_{A}(d)$ and $E_{V}(d)$ represent the total energies of the $A$ and the $V$ mode at the O$_2$-graphene distance $d$, respectively. The dotted lines represent Vinet function fits.}
 \label{fig:phase}
\end{figure}

The above results suggest a rotation of O$_2$ adsorbed at a hollow site from its parallel ($A$ or $B$) mode to the vertical $V$ mode when getting closer to the graphene surface.
Figure~\ref{fig:phase} presents the energy difference between the $A$ mode 
and the $V$ mode as a function of the O$_2$-graphene distance $d$. 
The DMC results show that a transition from the $A$ mode to the $V$ mode takes place around 2.8~\AA,
which seems to be reproduced, at least qualitatively, by DFT calculations based on various vdW-corrected functionals.
In the case of a DFT-PBE calculation without any vdW correction, however, such a transition is hardly noticeable, suggesting that it is mostly driven by the vdW interaction. 

\begin{figure}[t]
 \includegraphics[width=3 in]{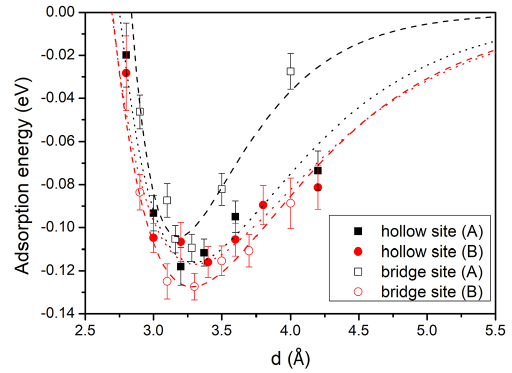}
 \caption{DMC adsorption energies of O$_2$ at the hollow and bridge sites for the $A$ and $B$ modes as functions of the distance from the graphene surface. The dotted lines indicate Morse function fits.}
 \label{fig:DMC_ad}
\end{figure}

For another possible adsorption site for O$_2$, we now consider a bridge site that was predicted in the DFT geometry optimization study of Guang~\cite{guang13} to be a stable adsorption site. For this only two parallel modes of $A$ and $B$ are investigated because the vertical $V$ mode was found to be less stable at the equilibrium distance than the parallel modes.
Figure~\ref{fig:DMC_ad} presents the DMC adsorption energies of O$_2$ at a bridge site, which were computed for the $2 \times 2 \times 1$ supercell, as functions of the distance from the graphene surface, along with the corresponding results at a hollow site.
Unlike an O$_2$ molecule adsorbed at a hollow site that showed no clear preference between $A$ and $B$ modes, O$_2$ at a bridge site is observed to favor the $B$ mode which shows a significantly lower equilibrium adsorption energy than the $A$ mode.
This suggests that an O$_2$ molecule at a bridge site is not allowed to rotate freely in the plane parallel to the graphene surface,
which can be understood by stronger repulsive interaction due to shorter O-C interatomic distance in the $A$ mode than in the $B$ mode.
Furthermore, our DMC calculations for the $2 \times 2 \times 1$ supercell reveal that the $B$-mode equilibrium adsorption energy at a bridge site is slightly lower than, or nearly identical to, the equilibrium adsorption energies at a hollow site.

\begin{figure}[t]
 \includegraphics[width=6 in]{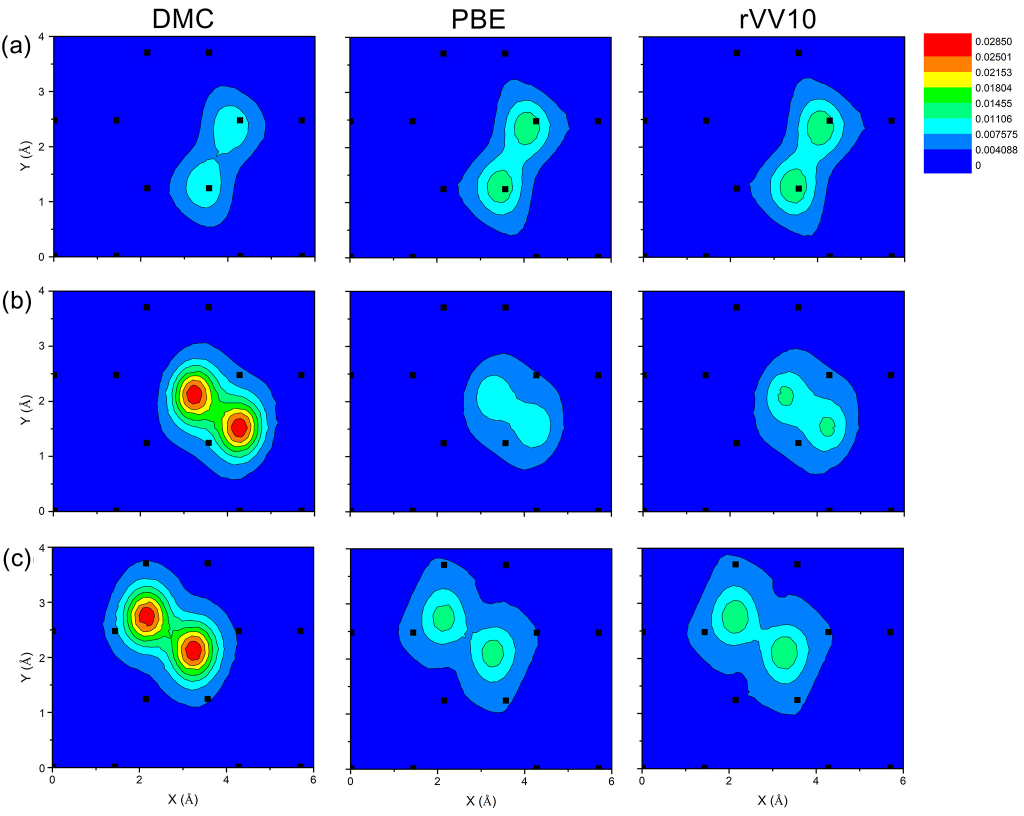}
 \caption{Two-dimensional charge density distributions $\rho_{inter}$ in the intermediate region between the graphene surface and O$_2$ for (a) the $A$-bridge, (b) the $B$-bridge, and (c) the $B$-hollow mode.  The results were obtained from the $2 \times 2 \times 1$ supercell calculations using DMC, PBE, and rVV10 methods. Black solid squares represent carbon atoms in the graphene sheet.}
 \label{fig:O2_contour}
\end{figure}

To understand the reason why the $B$ mode is preferred at a bridge site over the $A$ mode, we computed two-dimensional charge density distribution in the intermediate region between O$_2$ and the graphene surface:
\begin{equation}
   \rho_{inter}(x,y) = \rho(x,y, z_{gr}+r^{c}_{\rm C} < z < z_{{\rm O}_{2}}-r^{c}_{\rm O}), 
\end{equation}
where $z_{gr}$ and ${z_{\rm O_{2}}}$ denote $z$ coordinates of graphene and the oxygen dimer, respectively,
and $r^{c}_{\rm C}$ ($r^{c}_{\rm O}$) represents the covalent radius of a carbon (oxygen) atom.
Figure~\ref{fig:O2_contour} shows contour plots of $\rho_{inter}$ that were computed using DMC (left), PBE-DFT (middle) and rVV10-DFT (right) methods. 
One can see clear difference in the DMC density peaks between $A$-bridge and $B$-bridge modes.
Smaller amount of charges distributed in the intermediate region for the $A$ mode than for the $B$ mode suggests that electrons in the $A$-mode configuration tend to be pushed away from the intermediate region to reduce the repulsion due to the overlap between electron clouds of oxygen atoms and those of carbon atoms.
Note that oxygen atoms are closer to their nearest carbon atoms in the $A$-bridge mode than in the $B$-bridge mode.
This difference in the charge distributions between $A$ and $B$ modes is understood to account for the lower DMC adsorption energy of the $B$ mode at a bridge site. 
On the other hand, both PBE and rVV10 calculations do not show much difference in the amount of the intermediate charges between the $A$-bridge and the $B$-bridge modes, reflecting that the adsorption energies of these two orientation modes are similar to each other in both DFT calculations. 
From this we conclude that the inability of DFT to distinguish the adsorption energy between the two parallel orientation modes comes from its limitation in describing the electron-electron correlation, especially among the electrons in the intermediate region.
It can be also seen in Fig.~\ref{fig:O2_contour} that the DMC charge density distribution of the $B$-bridge mode do not show much difference from the corresponding result for the $B$-hollow mode, accounting for similar DMC adsorption energies between these two adsorption modes.

\begin{figure}[t]
 \includegraphics[width=6 in]{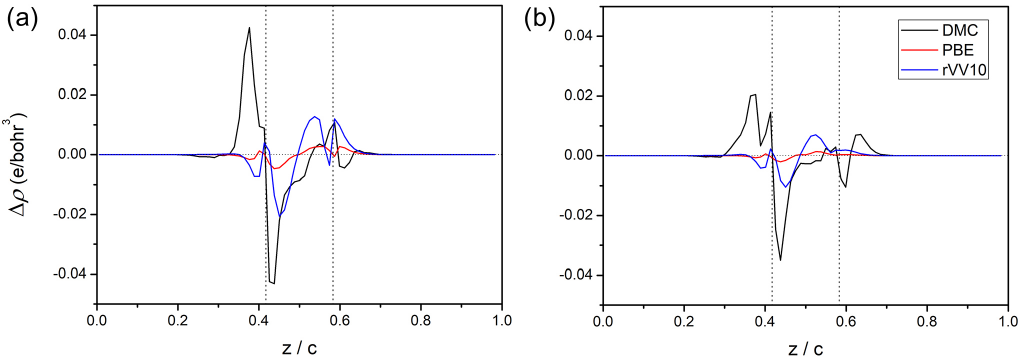}
 \caption{DMC and DFT charge density differences between the O$_2$-graphene complex and a system of a pristine graphene and an isolated O$_2$ molecule, which are projected to the vertical $z$ direction.
 The O$_2$ molecule in the O$_2$-graphene complex is adsorbed (a) at a hollow site and (b) at a bridge site while being in the $B$ orientation mode. 
 Left and right dotted vertical lines represent the locations of graphene and O$_2$, respectively.}
 \label{fig:1D_density}
\end{figure}

To understand different nature of the O$_2$ adsorption at a bridge site from the hollow-site adsorption, we now investigate the extent of charge redistribution induced by the graphene-O$_2$ interaction.
Figure~\ref{fig:1D_density} presents the difference between one-dimensional charge density distribution of the O$_2$-graphene complex and that of isolated graphene and O$_2$, which is projected to the vertical axis. 
It can be seen that the amounts of charges redistributed through the O$_2$ binding processes at both adsorption sites are larger in DMC calculations than in PBE and rVV10 calculations.
For both hollow- and bridge-site adsorptions, no noticeable charge accumulation is observed at the intermediate region between graphene and O$_2$, confirming the van der Waals nature of the O$_2$ binding to graphene. According to our DMC results, charges of graphene tend to be pushed to the opposite side of O$_2$ in both hollow- and bridge-site adsorptions.
A significant qualitative difference between DMC and DFT results is observed in the bridge-site adsorption, where the DMC result shows charges in O$_2$ being pushed away from the graphene side while both PBE and rVV10 calculations produce minimal charge redistribution on the O$_2$ side.
We understand that lack of charge redistribution around the bridge-site O$_2$ at the DFT level accounts for
no significant difference in the DFT adsorption energies between the $A$-bridge and the $B$-bridge modes.

\begin{table*}[t]
\centering
\caption{Extrapolated DMC adsorption energies $E_{ad}$ (eV) of O$_2$ at the equilibrium distances $d_{eq}$ (\AA) from a hollow and a bridge site, along with the corresponding DFT results based on several different exchange-correlation functionals.   
}
\label{tab:binding}
\begin{tabular}{ccccccccc}
\hline\hline
&  &  &  PBE & DFT-D2 & vdW-DF & vdW-DF2 & rVV10 & DMC  \\ \hline
\multirow{6}{*}{hollow} & \multirow{2}{*}{$A$ mode} & $d_{eq}$ &  3.63 &   3.17  &  3.28  &  3.14  & 3.10 & 3.28(5)  \\
                          &  & $E_{ad}$ &  -0.016  & -0.148 & -0.162 &  -0.145 &  -0.141 & -0.130(4) \\ \cline{2-9}
& \multirow{2}{*}{$B$ mode} & $d_{eq}$ & 3.63 & 3.17 & 3.28 & 3.14 & 3.10 & 3.30(5)  \\
                         &  & $E_{ad}$ & -0.016 & -0.148 & -0.164 & -0.146 & -0.142 & -0.126(4) \\ \cline{2-9}
& \multirow{2}{*}{$V$ mode} & $d_{eq}$ & 3.46 & 3.05 & 3.12 & 2.97 & 2.92 & 3.26(9) \\
                         &  & $E_{ad}$ & -0.014 & -0.121 & -0.130 & -0.116 & -0.110 & -0.111(7) \\ \hline
 \multirow{4}{*}{bridge} & \multirow{2}{*}{$A$ mode} & $d_{eq}$ & 3.66 & 3.22 & 3.32 &  3.19 &  3.15 & 3.18(3) \\
                         &  & $E_{ad}$ &  -0.017 & -0.141 & -0.158 &  -0.139 &  -0.132 & -0.117(7) \\ \cline{2-9} 
                         & \multirow{2}{*}{$B$ mode} & $d_{eq}$ &  3.71 & 3.23 & 3.33 &  3.21 &  3.16 & 3.21(2) \\
                         &  & $E_{ad}$ &  -0.016 & -0.135 & -0.156 &  -0.133 &  -0.130 & -0.135(4) \\ \hline \hline
\end{tabular}
\end{table*}

Finally we make finite size analysis of our DMC calculations to determine the most favorable adsorption site for O$_2$ along with its preferred orientation.
To reach the thermodynamic limit for each adsorption mode, we extrapolate DMC adsorption energies  computed at equilibrium distances for three different supercell sizes to the bulk limit, more details of which can be found in the Supplemental Material.~\footnote{See Supplemental Material at [URL will be
  inserted by publisher] for more information on finite-size analysis for the adsorption energy of O$_2$ on graphene.} 
Our extrapolated DMC results for equilibrium adsorption energies ($E_{ad}$) and the equilibrium distances ($d_{eq}$) are summarized in Table~\ref{tab:binding}, which also presents the corresponding DFT results computed with several different exchange-correlation functionals.
While DFT calculations with vdW-corrected functionals show clear binding of O$_2$ at both adsorption sites, the PBE calculation without a dispersion force predicts that an O$_2$ molecule is barely bound to either a hollow or a bridge site with its estimated binding energy less than 0.02 eV. This reflects the van der Waals nature of the O$_2$ binding to the graphene surface.
We note that vdW-DFT calculations tend to overestimate the O$_2$ binding energies but to underestimate its equilibrium distances when compared to the DMC results.
Among exchange-correlation functionals considered here, 
the O$_2$ adsorption energies are found to be best described, except for the $A$-bridge mode, with the rVV10 functional while vdW-DF (vdW-DF2) functional gives rise to the O$_2$-graphene equilibrium distances at a hollow (bridge) site that are most consistent with our DMC results.
No density functional is found to be able to recover simultaneously the DMC adsorption energies and equilibrium distances of O$_2$ on graphene. 
This limitation of the DFT calculations was also observed in recent theoretical studies for the interlayer binding of phosphorus~\cite{shulenburger15,ahn18} and carbon layers.~\cite{shin2017}

According to the results presented in Table~\ref{tab:binding},
regardless of the computational method used, two parallel modes of $A$ and $B$ have similar adsorption energies and equilibrium distances at a hollow site.
This suggests a free planar rotation of O$_2$ at a hollow site, which can be triggered by temperature or any external perturbation.
While all DFT calculations predict that these two parallel modes are nearly degenerate in energy even at a bridge site, our DMC results show that the $B$ mode is clearly favored over the $A$ mode for O$_2$ adsorbed at a bridge site.
This indicates that free planar rotation between $A$ and $B$ modes is no longer allowed for O$_2$ adsorbed at a bridge site, which is understood by stronger influence of repulsive interaction due to shorter O-C interatomic distance than for O$_2$ at a hollow site.
Furthermore, all DFT calculations presented in Table~\ref{tab:binding} favor the adsorption of O$_2$ at a hollow site over its adsorption at a bridge site, regardless of the orientation of O$_2$.
This is in contrast to our DMC result that the adsorption energy in the $B$-bridge mode is slightly lower than its hollow-site adsorption energies.

\begin{figure}[t]
 \includegraphics[width=6 in]{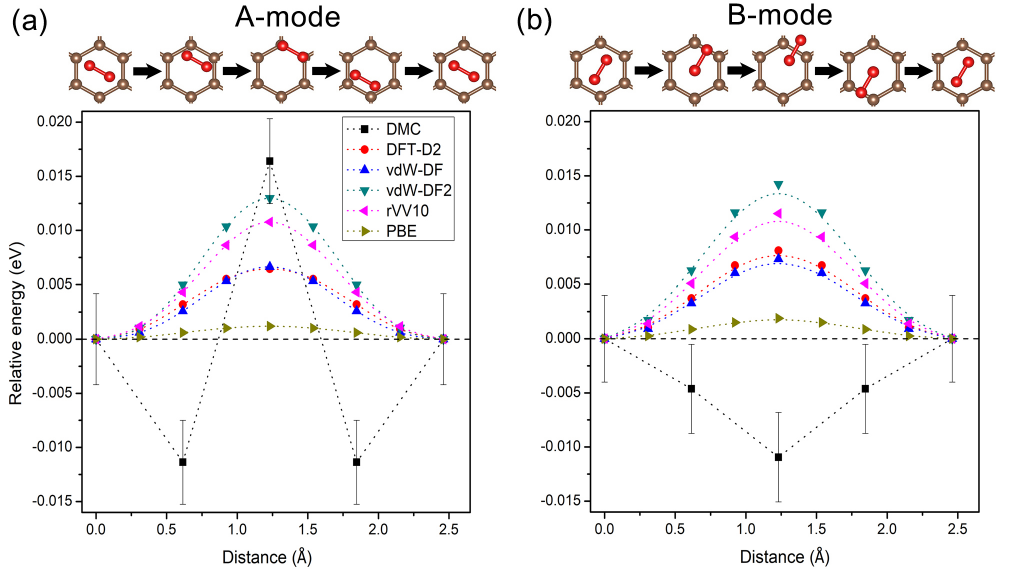}
 \caption{DMC and DFT in-plane diffusion barrier for the  hollow-bridge-hollow path for O$_2$ aligned in (a) $A$ and (b) $B$ orientation mode. Note that the adsorption energies at a hollow site are set to be zero and the dotted lines are just guides for the eyes.}
 \label{fig:O2_diffusion}
\end{figure}

The DMC adsorption energies presented in Table~\ref{tab:binding} are a little higher than the experimentally-reported value of -0.15 eV for the O$_2$ adsorption energy on graphene.~\cite{begsican17}
This leads us to explore the graphene surface for more favorable adsorption sites for O$_2$ than either a bridge or a hollow site and to investigate in the process diffusion mobility of O$_2$ on the graphene surface.
For this, the adsorption energy of O$_2$ is computed as it moves ''point by point" in the direction parallel to the graphene surface from a hollow site to a bridge site (see the top panel of Fig.~\ref{fig:O2_diffusion}) 
while the O$_2$-graphene distance is fixed to its hollow-site equilibrium distance.
The bottom panel of Fig.~\ref{fig:O2_diffusion} 
presents the O$_2$ adsorption energies along the diffusion path relative to the corresponding values at a hollow site. This reflects the above finding that the $B$ mode has a much lower DMC adsorption energy than the $A$ mode at a bridge site (note that the $A$ mode is nearly degenerate to the $B$ mode at a hollow site).
And it clearly shows the $B$-mode DMC adsorption energy is lower at a bridge site than at a hollow site.
Furthermore, our DMC results show that the $A$ orientation mode of O$_2$ has the lowest adsorption energy at an intermediate point between a hollow and a bridge site, where the adsorption energy is lower than, at least as low as, the adsorption energy of the $B$-bridge mode. 
The DMC $A$-mode adsorption energy at an intermediate site is estimated to be $-0.142(4)$ eV,
which is quantitatively consistent with the recently-reported experimental value.~\cite{begsican17}
According to our DMC results, either an intermediate point ($A$ mode) or a bridge site ($B$ mode), depending on the orientation mode, is energetically the most stable adsorption site for O$_2$.
On the other hand, all DFT results presented in Fig.~\ref{fig:O2_diffusion} show that the adsorption energy of O$_2$ increases steadily as it moves from a hollow site to a bridge site,
indicating that a bridge site is the peak point in the diffusion path for O$_2$ on the graphene surface. 
This contradicts our DMC prediction that a hollow site is the peak point in the O$_2$ diffusion path.
As far as the diffusion barrier of O$_2$ along a bridge-hollow-bridge diffusion path is concerned, our DMC estimate is 11(4) meV, similar to its rVV10 and vdW-DF2 estimates.

\section{\label{sec:summary} Conclusions}
We have investigated the relative stability of three orientation modes of O$_2$ adsorbed on a single layer graphene using DMC and vdW-corrected DFT calculations. 
While the two orientation modes parallel to the graphene surface are found to be more favorable than the vertical mode at the equilibrium adsorption distances, the vertical mode is favored at shorter distances where strong repulsive force between graphene and O$_2$ is more dominant. 
While vdW-corrected functionals were able to describe, at least qualitatively, the orientation profile of O$_2$ adsorbed on graphene, no functional was able to reproduce simultaneously the equilibrium adsorption distance and adsorption energy estimated in the DMC calculations.

Similar adsorption energies for its two parallel orientation modes of $A$ and $B$ led us to predict a nearly free planar rotation of O$_2$ adsorbed at a hollow site.
On the other hand, our DMC calculations for the O$_2$ adsorption at a bridge site show that the $B$ mode has a significantly lower adsorption energy than the $A$ mode,
breaking the rotational symmetry between these two modes.
Further DMC investigation of the O$_2$ adsorption reveals that in contrast to DFT calculations, a hollow site is not the most stable adsorption site for O$_2$.
The site with the lowest O$_2$ adsorption energy was found to be a bridge site for the $B$ orientation mode and an intermediate site between a hollow and a bridge site for the $A$ mode.
The stable $A$-mode adsorption of O$_2$ at an intermediate site can be attributed to the interplay between repulsive interaction and vdW interaction.
As an O$_2$ molecule in the $A$-mode orientation approaches a bridge site from a hollow site, an attractive interaction due to dispersion force gets stronger but the strength of the repulsive interaction due to charge overlap between O and C atoms also increases, resulting in the lowest adsorption energy at an intermediate site. 

Unlike DFT results based on several different exchange-correlation functionals, our DMC study predicts that the $B$-bridge mode is one of the stable adsorption modes. 
Based on low diffusion barrier of $\sim 11$ meV, we conclude that the adsorbed O$_2$ molecule is very diffusive on the graphene surface. 
The most stable O$_2$ adsorption site is concluded to be degenerate between an intermediate site and a bridge site, depending on the molecular orientation of O$_2$.
This can be hardly reproduced within DFT framework because the O$_2$-graphene interaction in the $B$-bridge mode is not accurately described even with vdW-corrected exchange-correlation functionals.
Finally we expect that our DMC density distributions can provide a guideline to find better exchange-correlation functionals for accurate theoretical estimations of gas sensing properties of graphene, possibly other 2D materials.    

\begin{acknowledgements}
H.S, Y.L and A.B were supported by the U.S. Department of Energy, Office of Science, Basic Energy Sciences, Materials Sciences and Engineering Division, as part of the Computational Materials Sciences Program and Center for Predictive Simulation of Functional Materials.
Y.K was supported by the Basic Science Research Program (2018R1D1A1B07042443) through the National Research Foundation of Korea funded by the Ministry of Education.
An award of computer time was provided by the Innovative and Novel Computational Impact on Theory and Experiment (INCITE) program. This research used resources of the Argonne Leadership Computing Facility, which is a DOE Office of Science User Facility supported under contract DE-AC02-06CH11357. 
\end{acknowledgements}

\bibliography{all}

\begin{thebibliography}{60}%
\makeatletter
\providecommand \@ifxundefined [1]{%
 \@ifx{#1\undefined}
}%
\providecommand \@ifnum [1]{%
 \ifnum #1\expandafter \@firstoftwo
 \else \expandafter \@secondoftwo
 \fi
}%
\providecommand \@ifx [1]{%
 \ifx #1\expandafter \@firstoftwo
 \else \expandafter \@secondoftwo
 \fi
}%
\providecommand \natexlab [1]{#1}%
\providecommand \enquote  [1]{``#1''}%
\providecommand \bibnamefont  [1]{#1}%
\providecommand \bibfnamefont [1]{#1}%
\providecommand \citenamefont [1]{#1}%
\providecommand \href@noop [0]{\@secondoftwo}%
\providecommand \href [0]{\begingroup \@sanitize@url \@href}%
\providecommand \@href[1]{\@@startlink{#1}\@@href}%
\providecommand \@@href[1]{\endgroup#1\@@endlink}%
\providecommand \@sanitize@url [0]{\catcode `\\12\catcode `\$12\catcode
  `\&12\catcode `\#12\catcode `\^12\catcode `\_12\catcode `\%12\relax}%
\providecommand \@@startlink[1]{}%
\providecommand \@@endlink[0]{}%
\providecommand \url  [0]{\begingroup\@sanitize@url \@url }%
\providecommand \@url [1]{\endgroup\@href {#1}{\urlprefix }}%
\providecommand \urlprefix  [0]{URL }%
\providecommand \Eprint [0]{\href }%
\providecommand \doibase [0]{http://dx.doi.org/}%
\providecommand \selectlanguage [0]{\@gobble}%
\providecommand \bibinfo  [0]{\@secondoftwo}%
\providecommand \bibfield  [0]{\@secondoftwo}%
\providecommand \translation [1]{[#1]}%
\providecommand \BibitemOpen [0]{}%
\providecommand \bibitemStop [0]{}%
\providecommand \bibitemNoStop [0]{.\EOS\space}%
\providecommand \EOS [0]{\spacefactor3000\relax}%
\providecommand \BibitemShut  [1]{\csname bibitem#1\endcsname}%
\let\auto@bib@innerbib\@empty
\bibitem [{\citenamefont {Novoselov}\ \emph {et~al.}(2005)\citenamefont
  {Novoselov}, \citenamefont {Geim}, \citenamefont {Morozov}, \citenamefont
  {Jiang}, \citenamefont {Katsnelson}, \citenamefont {Grigorieva},
  \citenamefont {Dubonos},\ and\ \citenamefont {Frisov}}]{novoselov05}%
  \BibitemOpen
  \bibfield  {author} {\bibinfo {author} {\bibfnamefont {K.~S.}\ \bibnamefont
  {Novoselov}}, \bibinfo {author} {\bibfnamefont {A.~K.}\ \bibnamefont {Geim}},
  \bibinfo {author} {\bibfnamefont {S.~V.}\ \bibnamefont {Morozov}}, \bibinfo
  {author} {\bibfnamefont {D.}~\bibnamefont {Jiang}}, \bibinfo {author}
  {\bibfnamefont {M.~I.}\ \bibnamefont {Katsnelson}}, \bibinfo {author}
  {\bibfnamefont {I.~V.}\ \bibnamefont {Grigorieva}}, \bibinfo {author}
  {\bibfnamefont {S.~V.}\ \bibnamefont {Dubonos}}, \ and\ \bibinfo {author}
  {\bibfnamefont {A.~A.}\ \bibnamefont {Frisov}},\ }\href {\doibase
  10.1038/nature04233} {\bibfield  {journal} {\bibinfo  {journal} {Nature}\
  }\textbf {\bibinfo {volume} {438}},\ \bibinfo {pages} {197} (\bibinfo {year}
  {2005})}\BibitemShut {NoStop}%
\bibitem [{\citenamefont {Zhang}\ \emph {et~al.}(2005)\citenamefont {Zhang},
  \citenamefont {Tan}, \citenamefont {Stormer},\ and\ \citenamefont
  {Kim}}]{zhang05}%
  \BibitemOpen
  \bibfield  {author} {\bibinfo {author} {\bibfnamefont {Y.}~\bibnamefont
  {Zhang}}, \bibinfo {author} {\bibfnamefont {Y.-W.}\ \bibnamefont {Tan}},
  \bibinfo {author} {\bibfnamefont {H.~L.}\ \bibnamefont {Stormer}}, \ and\
  \bibinfo {author} {\bibfnamefont {P.}~\bibnamefont {Kim}},\ }\href {\doibase
  10.1038/nature04235} {\bibfield  {journal} {\bibinfo  {journal} {Nature}\
  }\textbf {\bibinfo {volume} {438}},\ \bibinfo {pages} {201} (\bibinfo {year}
  {2005})}\BibitemShut {NoStop}%
\bibitem [{\citenamefont {Geim}\ and\ \citenamefont
  {Novoselov}(2007)}]{geim07}%
  \BibitemOpen
  \bibfield  {author} {\bibinfo {author} {\bibfnamefont {A.~K.}\ \bibnamefont
  {Geim}}\ and\ \bibinfo {author} {\bibfnamefont {K.~S.}\ \bibnamefont
  {Novoselov}},\ }\href {\doibase 10.1038/nmat1849} {\bibfield  {journal}
  {\bibinfo  {journal} {Nat.\ Mater}\ }\textbf {\bibinfo {volume} {6}},\
  \bibinfo {pages} {183} (\bibinfo {year} {2007})}\BibitemShut {NoStop}%
\bibitem [{\citenamefont {Geim}(2009)}]{geim09}%
  \BibitemOpen
  \bibfield  {author} {\bibinfo {author} {\bibfnamefont {A.~K.}\ \bibnamefont
  {Geim}},\ }\href {\doibase 10.1126/science.1158877} {\bibfield  {journal}
  {\bibinfo  {journal} {Science}\ }\textbf {\bibinfo {volume} {324}},\ \bibinfo
  {pages} {1530} (\bibinfo {year} {2009})}\BibitemShut {NoStop}%
\bibitem [{\citenamefont {Kong}\ \emph {et~al.}(2000)\citenamefont {Kong},
  \citenamefont {Franklin}, \citenamefont {Zhou}, \citenamefont {Chapline},
  \citenamefont {Peng}, \citenamefont {Cho},\ and\ \citenamefont
  {Dai}}]{kong00}%
  \BibitemOpen
  \bibfield  {author} {\bibinfo {author} {\bibfnamefont {J.}~\bibnamefont
  {Kong}}, \bibinfo {author} {\bibfnamefont {N.~R.}\ \bibnamefont {Franklin}},
  \bibinfo {author} {\bibfnamefont {C.}~\bibnamefont {Zhou}}, \bibinfo {author}
  {\bibfnamefont {M.~G.}\ \bibnamefont {Chapline}}, \bibinfo {author}
  {\bibfnamefont {S.}~\bibnamefont {Peng}}, \bibinfo {author} {\bibfnamefont
  {K.}~\bibnamefont {Cho}}, \ and\ \bibinfo {author} {\bibfnamefont
  {H.}~\bibnamefont {Dai}},\ }\href@noop {} {\bibfield  {journal} {\bibinfo
  {journal} {Science}\ }\textbf {\bibinfo {volume} {287}},\ \bibinfo {pages}
  {622} (\bibinfo {year} {2000})}\BibitemShut {NoStop}%
\bibitem [{\citenamefont {Cui}\ \emph {et~al.}(2001)\citenamefont {Cui},
  \citenamefont {Wei}, \citenamefont {Park},\ and\ \citenamefont
  {Lieber}}]{cui01}%
  \BibitemOpen
  \bibfield  {author} {\bibinfo {author} {\bibfnamefont {Y.}~\bibnamefont
  {Cui}}, \bibinfo {author} {\bibfnamefont {Q.}~\bibnamefont {Wei}}, \bibinfo
  {author} {\bibfnamefont {H.}~\bibnamefont {Park}}, \ and\ \bibinfo {author}
  {\bibfnamefont {C.~M.}\ \bibnamefont {Lieber}},\ }\href@noop {} {\bibfield
  {journal} {\bibinfo  {journal} {Science}\ }\textbf {\bibinfo {volume}
  {293}},\ \bibinfo {pages} {1289} (\bibinfo {year} {2001})}\BibitemShut
  {NoStop}%
\bibitem [{\citenamefont {Schedin}\ \emph {et~al.}(2007)\citenamefont
  {Schedin}, \citenamefont {Geim}, \citenamefont {Morozov}, \citenamefont
  {Hill}, \citenamefont {Blake}, \citenamefont {Katsnelson},\ and\
  \citenamefont {Novoselov}}]{schedin07}%
  \BibitemOpen
  \bibfield  {author} {\bibinfo {author} {\bibfnamefont {F.}~\bibnamefont
  {Schedin}}, \bibinfo {author} {\bibfnamefont {A.~K.}\ \bibnamefont {Geim}},
  \bibinfo {author} {\bibfnamefont {S.~V.}\ \bibnamefont {Morozov}}, \bibinfo
  {author} {\bibfnamefont {E.~W.}\ \bibnamefont {Hill}}, \bibinfo {author}
  {\bibfnamefont {P.}~\bibnamefont {Blake}}, \bibinfo {author} {\bibfnamefont
  {M.~I.}\ \bibnamefont {Katsnelson}}, \ and\ \bibinfo {author} {\bibfnamefont
  {K.~S.}\ \bibnamefont {Novoselov}},\ }\href {\doibase 10.1038/nmat1967}
  {\bibfield  {journal} {\bibinfo  {journal} {Nat. Mater}\ }\textbf {\bibinfo
  {volume} {6}},\ \bibinfo {pages} {652} (\bibinfo {year} {2007})}\BibitemShut
  {NoStop}%
\bibitem [{\citenamefont {Lu}\ \emph {et~al.}(2009)\citenamefont {Lu},
  \citenamefont {Ocola},\ and\ \citenamefont {Chen}}]{lu09}%
  \BibitemOpen
  \bibfield  {author} {\bibinfo {author} {\bibfnamefont {G.}~\bibnamefont
  {Lu}}, \bibinfo {author} {\bibfnamefont {L.~E.}\ \bibnamefont {Ocola}}, \
  and\ \bibinfo {author} {\bibfnamefont {J.}~\bibnamefont {Chen}},\ }\href@noop
  {} {\bibfield  {journal} {\bibinfo  {journal} {Nanotechnology}\ }\textbf
  {\bibinfo {volume} {20}},\ \bibinfo {pages} {445502} (\bibinfo {year}
  {2009})}\BibitemShut {NoStop}%
\bibitem [{\citenamefont {Varghese}\ \emph {et~al.}(2015)\citenamefont
  {Varghese}, \citenamefont {Lonkar}, \citenamefont {Singh}, \citenamefont
  {Swaminathan},\ and\ \citenamefont {Abdala}}]{varghese15}%
  \BibitemOpen
  \bibfield  {author} {\bibinfo {author} {\bibfnamefont {S.~S.}\ \bibnamefont
  {Varghese}}, \bibinfo {author} {\bibfnamefont {S.}~\bibnamefont {Lonkar}},
  \bibinfo {author} {\bibfnamefont {K.~K.}\ \bibnamefont {Singh}}, \bibinfo
  {author} {\bibfnamefont {S.}~\bibnamefont {Swaminathan}}, \ and\ \bibinfo
  {author} {\bibfnamefont {A.}~\bibnamefont {Abdala}},\ }\href@noop {}
  {\bibfield  {journal} {\bibinfo  {journal} {Sens. Actuators B}\ }\textbf
  {\bibinfo {volume} {218}},\ \bibinfo {pages} {160} (\bibinfo {year}
  {2015})}\BibitemShut {NoStop}%
\bibitem [{\citenamefont {Wehling}\ \emph {et~al.}(2008)\citenamefont
  {Wehling}, \citenamefont {Novoselov}, \citenamefont {Morozov}, \citenamefont
  {Vdovin}, \citenamefont {Katsnelson}, \citenamefont {Geim},\ and\
  \citenamefont {Lichtenstein}}]{wehling08}%
  \BibitemOpen
  \bibfield  {author} {\bibinfo {author} {\bibfnamefont {T.~O.}\ \bibnamefont
  {Wehling}}, \bibinfo {author} {\bibfnamefont {K.~S.}\ \bibnamefont
  {Novoselov}}, \bibinfo {author} {\bibfnamefont {S.~V.}\ \bibnamefont
  {Morozov}}, \bibinfo {author} {\bibfnamefont {E.~E.}\ \bibnamefont {Vdovin}},
  \bibinfo {author} {\bibfnamefont {M.~I.}\ \bibnamefont {Katsnelson}},
  \bibinfo {author} {\bibfnamefont {A.~K.}\ \bibnamefont {Geim}}, \ and\
  \bibinfo {author} {\bibfnamefont {A.~I.}\ \bibnamefont {Lichtenstein}},\
  }\href@noop {} {\bibfield  {journal} {\bibinfo  {journal} {Nano Lett.}\
  }\textbf {\bibinfo {volume} {8}},\ \bibinfo {pages} {173} (\bibinfo {year}
  {2008})}\BibitemShut {NoStop}%
\bibitem [{\citenamefont {Lee}\ and\ \citenamefont {Kim}(2013)}]{lee13}%
  \BibitemOpen
  \bibfield  {author} {\bibinfo {author} {\bibfnamefont {K.}~\bibnamefont
  {Lee}}\ and\ \bibinfo {author} {\bibfnamefont {S.}~\bibnamefont {Kim}},\
  }\href@noop {} {\bibfield  {journal} {\bibinfo  {journal} {Bull. Korean Chem.
  Soc.}\ }\textbf {\bibinfo {volume} {34}},\ \bibinfo {pages} {3022} (\bibinfo
  {year} {2013})}\BibitemShut {NoStop}%
\bibitem [{\citenamefont {Chen}\ \emph {et~al.}(2014)\citenamefont {Chen},
  \citenamefont {Darancet}, \citenamefont {Wang}, \citenamefont {Crowther},
  \citenamefont {Gao}, \citenamefont {Dean}, \citenamefont {Taniguchi},
  \citenamefont {Watanabe}, \citenamefont {Hone}, \citenamefont {Marianetti},\
  and\ \citenamefont {Brus}}]{chen14}%
  \BibitemOpen
  \bibfield  {author} {\bibinfo {author} {\bibfnamefont {Z.}~\bibnamefont
  {Chen}}, \bibinfo {author} {\bibfnamefont {P.}~\bibnamefont {Darancet}},
  \bibinfo {author} {\bibfnamefont {L.}~\bibnamefont {Wang}}, \bibinfo {author}
  {\bibfnamefont {A.~C.}\ \bibnamefont {Crowther}}, \bibinfo {author}
  {\bibfnamefont {Y.}~\bibnamefont {Gao}}, \bibinfo {author} {\bibfnamefont
  {C.~R.}\ \bibnamefont {Dean}}, \bibinfo {author} {\bibfnamefont
  {T.}~\bibnamefont {Taniguchi}}, \bibinfo {author} {\bibfnamefont
  {K.}~\bibnamefont {Watanabe}}, \bibinfo {author} {\bibfnamefont
  {J.}~\bibnamefont {Hone}}, \bibinfo {author} {\bibfnamefont {C.~A.}\
  \bibnamefont {Marianetti}}, \ and\ \bibinfo {author} {\bibfnamefont {L.~E.}\
  \bibnamefont {Brus}},\ }\href@noop {} {\bibfield  {journal} {\bibinfo
  {journal} {ACS Nano}\ }\textbf {\bibinfo {volume} {8}},\ \bibinfo {pages}
  {2943} (\bibinfo {year} {2014})}\BibitemShut {NoStop}%
\bibitem [{\citenamefont {Ganji}\ \emph {et~al.}(2015)\citenamefont {Ganji},
  \citenamefont {Hosseini-Khah},\ and\ \citenamefont {Amini-Tabar}}]{ganji15}%
  \BibitemOpen
  \bibfield  {author} {\bibinfo {author} {\bibfnamefont {M.~D.}\ \bibnamefont
  {Ganji}}, \bibinfo {author} {\bibfnamefont {S.~M.}\ \bibnamefont
  {Hosseini-Khah}}, \ and\ \bibinfo {author} {\bibfnamefont {Z.}~\bibnamefont
  {Amini-Tabar}},\ }\href@noop {} {\bibfield  {journal} {\bibinfo  {journal}
  {Phys. Chem. Chem. Phys.}\ }\textbf {\bibinfo {volume} {17}},\ \bibinfo
  {pages} {2504} (\bibinfo {year} {2015})}\BibitemShut {NoStop}%
\bibitem [{\citenamefont {Ratinac}\ \emph {et~al.}(2010)\citenamefont
  {Ratinac}, \citenamefont {Yang}, \citenamefont {Ringer},\ and\ \citenamefont
  {Braet}}]{ratinac10}%
  \BibitemOpen
  \bibfield  {author} {\bibinfo {author} {\bibfnamefont {K.~R.}\ \bibnamefont
  {Ratinac}}, \bibinfo {author} {\bibfnamefont {W.}~\bibnamefont {Yang}},
  \bibinfo {author} {\bibfnamefont {S.~P.}\ \bibnamefont {Ringer}}, \ and\
  \bibinfo {author} {\bibfnamefont {F.}~\bibnamefont {Braet}},\ }\href
  {\doibase 10.1021/es902659d} {\bibfield  {journal} {\bibinfo  {journal}
  {Environ. Sci. Technol.}\ }\textbf {\bibinfo {volume} {44}},\ \bibinfo
  {pages} {1167} (\bibinfo {year} {2010})}\BibitemShut {NoStop}%
\bibitem [{\citenamefont {Yuan}\ and\ \citenamefont {Shi}(2013)}]{yuan13}%
  \BibitemOpen
  \bibfield  {author} {\bibinfo {author} {\bibfnamefont {W.}~\bibnamefont
  {Yuan}}\ and\ \bibinfo {author} {\bibfnamefont {G.}~\bibnamefont {Shi}},\
  }\href {\doibase 10.1039/c3ta11774j} {\bibfield  {journal} {\bibinfo
  {journal} {J. Mater. Chem. A}\ }\textbf {\bibinfo {volume} {1}},\ \bibinfo
  {pages} {10078} (\bibinfo {year} {2013})}\BibitemShut {NoStop}%
\bibitem [{\citenamefont {Jaaniso}\ \emph {et~al.}(2014)\citenamefont
  {Jaaniso}, \citenamefont {Kahro}, \citenamefont {Kozlova}, \citenamefont
  {Aarik}, \citenamefont {Aarik}, \citenamefont {Alles}, \citenamefont
  {Floren}, \citenamefont {Gerst}, \citenamefont {Kasikov}, \citenamefont
  {Niilisk},\ and\ \citenamefont {Sammelselg}}]{jaaniso14}%
  \BibitemOpen
  \bibfield  {author} {\bibinfo {author} {\bibfnamefont {R.}~\bibnamefont
  {Jaaniso}}, \bibinfo {author} {\bibfnamefont {T.}~\bibnamefont {Kahro}},
  \bibinfo {author} {\bibfnamefont {J.}~\bibnamefont {Kozlova}}, \bibinfo
  {author} {\bibfnamefont {J.}~\bibnamefont {Aarik}}, \bibinfo {author}
  {\bibfnamefont {L.}~\bibnamefont {Aarik}}, \bibinfo {author} {\bibfnamefont
  {H.}~\bibnamefont {Alles}}, \bibinfo {author} {\bibfnamefont
  {A.}~\bibnamefont {Floren}}, \bibinfo {author} {\bibfnamefont
  {A.}~\bibnamefont {Gerst}}, \bibinfo {author} {\bibfnamefont
  {A.}~\bibnamefont {Kasikov}}, \bibinfo {author} {\bibfnamefont
  {A.}~\bibnamefont {Niilisk}}, \ and\ \bibinfo {author} {\bibfnamefont
  {V.}~\bibnamefont {Sammelselg}},\ }\href {\doibase 10.1016/j.snb.2013.09.068}
  {\bibfield  {journal} {\bibinfo  {journal} {Sens. Actuators B}\ }\textbf
  {\bibinfo {volume} {190}},\ \bibinfo {pages} {1006} (\bibinfo {year}
  {2014})}\BibitemShut {NoStop}%
\bibitem [{\citenamefont {Novikov}\ \emph {et~al.}(2016)\citenamefont
  {Novikov}, \citenamefont {Lebedeva}, \citenamefont {Satrapinski},
  \citenamefont {Walden}, \citenamefont {Davydov},\ and\ \citenamefont
  {Lebedev}}]{novikov16}%
  \BibitemOpen
  \bibfield  {author} {\bibinfo {author} {\bibfnamefont {S.}~\bibnamefont
  {Novikov}}, \bibinfo {author} {\bibfnamefont {N.}~\bibnamefont {Lebedeva}},
  \bibinfo {author} {\bibfnamefont {A.}~\bibnamefont {Satrapinski}}, \bibinfo
  {author} {\bibfnamefont {J.}~\bibnamefont {Walden}}, \bibinfo {author}
  {\bibfnamefont {V.}~\bibnamefont {Davydov}}, \ and\ \bibinfo {author}
  {\bibfnamefont {A.}~\bibnamefont {Lebedev}},\ }\href {\doibase
  10.1016/j.snb.2016.05.114} {\bibfield  {journal} {\bibinfo  {journal} {Sens.
  Actuators B}\ }\textbf {\bibinfo {volume} {236}},\ \bibinfo {pages} {1054}
  (\bibinfo {year} {2016})}\BibitemShut {NoStop}%
\bibitem [{\citenamefont {Huang}\ \emph {et~al.}(2008)\citenamefont {Huang},
  \citenamefont {Li}, \citenamefont {Liu}, \citenamefont {Zhou}, \citenamefont
  {Hao}, \citenamefont {Wu}, \citenamefont {Gu},\ and\ \citenamefont
  {Duan}}]{huang08}%
  \BibitemOpen
  \bibfield  {author} {\bibinfo {author} {\bibfnamefont {B.}~\bibnamefont
  {Huang}}, \bibinfo {author} {\bibfnamefont {Z.}~\bibnamefont {Li}}, \bibinfo
  {author} {\bibfnamefont {Z.}~\bibnamefont {Liu}}, \bibinfo {author}
  {\bibfnamefont {G.}~\bibnamefont {Zhou}}, \bibinfo {author} {\bibfnamefont
  {S.}~\bibnamefont {Hao}}, \bibinfo {author} {\bibfnamefont {J.}~\bibnamefont
  {Wu}}, \bibinfo {author} {\bibfnamefont {B.-L.}\ \bibnamefont {Gu}}, \ and\
  \bibinfo {author} {\bibfnamefont {W.}~\bibnamefont {Duan}},\ }\href {\doibase
  10.1021/jp8021024} {\bibfield  {journal} {\bibinfo  {journal} {J. Phys. Chem.
  C}\ }\textbf {\bibinfo {volume} {112}},\ \bibinfo {pages} {13442} (\bibinfo
  {year} {2008})}\BibitemShut {NoStop}%
\bibitem [{\citenamefont {Leenaerts}\ \emph {et~al.}(2009)\citenamefont
  {Leenaerts}, \citenamefont {Partoens},\ and\ \citenamefont
  {Peeters}}]{leenaerts09}%
  \BibitemOpen
  \bibfield  {author} {\bibinfo {author} {\bibfnamefont {O.}~\bibnamefont
  {Leenaerts}}, \bibinfo {author} {\bibfnamefont {B.}~\bibnamefont {Partoens}},
  \ and\ \bibinfo {author} {\bibfnamefont {F.~M.}\ \bibnamefont {Peeters}},\
  }\href {\doibase 10.1016/j.mejo.2008.11.022} {\bibfield  {journal} {\bibinfo
  {journal} {Microelectron. J.}\ }\textbf {\bibinfo {volume} {40}},\ \bibinfo
  {pages} {860} (\bibinfo {year} {2009})}\BibitemShut {NoStop}%
\bibitem [{\citenamefont {Zhou}\ \emph {et~al.}(2011)\citenamefont {Zhou},
  \citenamefont {Lu}, \citenamefont {Cai}, \citenamefont {Zhang},\ and\
  \citenamefont {Feng}}]{zhou11}%
  \BibitemOpen
  \bibfield  {author} {\bibinfo {author} {\bibfnamefont {M.}~\bibnamefont
  {Zhou}}, \bibinfo {author} {\bibfnamefont {Y.-H.}\ \bibnamefont {Lu}},
  \bibinfo {author} {\bibfnamefont {Y.-Q.}\ \bibnamefont {Cai}}, \bibinfo
  {author} {\bibfnamefont {C.}~\bibnamefont {Zhang}}, \ and\ \bibinfo {author}
  {\bibfnamefont {Y.-P.}\ \bibnamefont {Feng}},\ }\href {\doibase
  10.1088/0957-4484/22/38/385502} {\bibfield  {journal} {\bibinfo  {journal}
  {Nanotechnology}\ }\textbf {\bibinfo {volume} {22}},\ \bibinfo {pages}
  {385502} (\bibinfo {year} {2011})}\BibitemShut {NoStop}%
\bibitem [{\citenamefont {Choudhuri}\ \emph {et~al.}(2015)\citenamefont
  {Choudhuri}, \citenamefont {Patra}, \citenamefont {Mahata}, \citenamefont
  {Ahuja},\ and\ \citenamefont {Pathak}}]{choudhuri15}%
  \BibitemOpen
  \bibfield  {author} {\bibinfo {author} {\bibfnamefont {I.}~\bibnamefont
  {Choudhuri}}, \bibinfo {author} {\bibfnamefont {N.}~\bibnamefont {Patra}},
  \bibinfo {author} {\bibfnamefont {A.}~\bibnamefont {Mahata}}, \bibinfo
  {author} {\bibfnamefont {R.}~\bibnamefont {Ahuja}}, \ and\ \bibinfo {author}
  {\bibfnamefont {B.}~\bibnamefont {Pathak}},\ }\href {\doibase
  10.1021/acs.jpcc.5b07359} {\bibfield  {journal} {\bibinfo  {journal} {J.
  Phys. Chem. C}\ }\textbf {\bibinfo {volume} {119}},\ \bibinfo {pages} {24827}
  (\bibinfo {year} {2015})}\BibitemShut {NoStop}%
\bibitem [{\citenamefont {Liang}\ \emph {et~al.}(2017)\citenamefont {Liang},
  \citenamefont {Ding}, \citenamefont {Ng},\ and\ \citenamefont
  {Wu}}]{liang17}%
  \BibitemOpen
  \bibfield  {author} {\bibinfo {author} {\bibfnamefont {X.-Y.}\ \bibnamefont
  {Liang}}, \bibinfo {author} {\bibfnamefont {N.}~\bibnamefont {Ding}},
  \bibinfo {author} {\bibfnamefont {S.-P.}\ \bibnamefont {Ng}}, \ and\ \bibinfo
  {author} {\bibfnamefont {C.-M.~L.}\ \bibnamefont {Wu}},\ }\href {\doibase
  10.1016/j.apsusc.2017.03.178} {\bibfield  {journal} {\bibinfo  {journal}
  {Appl. Surf. Sci.}\ }\textbf {\bibinfo {volume} {411}},\ \bibinfo {pages}
  {11} (\bibinfo {year} {2017})}\BibitemShut {NoStop}%
\bibitem [{\citenamefont {Kauffman}\ \emph {et~al.}(2009)\citenamefont
  {Kauffman}, \citenamefont {Shade}, \citenamefont {Oh}, \citenamefont
  {Petoud},\ and\ \citenamefont {Star}}]{kauffman09}%
  \BibitemOpen
  \bibfield  {author} {\bibinfo {author} {\bibfnamefont {D.~R.}\ \bibnamefont
  {Kauffman}}, \bibinfo {author} {\bibfnamefont {C.~M.}\ \bibnamefont {Shade}},
  \bibinfo {author} {\bibfnamefont {H.}~\bibnamefont {Oh}}, \bibinfo {author}
  {\bibfnamefont {S.}~\bibnamefont {Petoud}}, \ and\ \bibinfo {author}
  {\bibfnamefont {A.}~\bibnamefont {Star}},\ }\href {\doibase
  10.1038/nchem.323} {\bibfield  {journal} {\bibinfo  {journal} {Nat. Chem.}\
  }\textbf {\bibinfo {volume} {1}},\ \bibinfo {pages} {500} (\bibinfo {year}
  {2009})}\BibitemShut {NoStop}%
\bibitem [{\citenamefont {Rajavel}\ \emph {et~al.}(2015)\citenamefont
  {Rajavel}, \citenamefont {Lalitha}, \citenamefont {Radhakrishnan},
  \citenamefont {Senthilkumar},\ and\ \citenamefont {Kumar}}]{rajavel15}%
  \BibitemOpen
  \bibfield  {author} {\bibinfo {author} {\bibfnamefont {K.}~\bibnamefont
  {Rajavel}}, \bibinfo {author} {\bibfnamefont {M.}~\bibnamefont {Lalitha}},
  \bibinfo {author} {\bibfnamefont {J.~K.}\ \bibnamefont {Radhakrishnan}},
  \bibinfo {author} {\bibfnamefont {L.}~\bibnamefont {Senthilkumar}}, \ and\
  \bibinfo {author} {\bibfnamefont {R.~T.~R.}\ \bibnamefont {Kumar}},\
  }\href@noop {} {\bibfield  {journal} {\bibinfo  {journal} {ACS Appl. Mater.
  Interfaces}\ }\textbf {\bibinfo {volume} {7}},\ \bibinfo {pages} {23857}
  (\bibinfo {year} {2015})}\BibitemShut {NoStop}%
\bibitem [{\citenamefont {Tang}\ \emph {et~al.}(2017)\citenamefont {Tang},
  \citenamefont {Wu}, \citenamefont {Liu},\ and\ \citenamefont {Gu}}]{tang17}%
  \BibitemOpen
  \bibfield  {author} {\bibinfo {author} {\bibfnamefont {S.}~\bibnamefont
  {Tang}}, \bibinfo {author} {\bibfnamefont {W.}~\bibnamefont {Wu}}, \bibinfo
  {author} {\bibfnamefont {L.}~\bibnamefont {Liu}}, \ and\ \bibinfo {author}
  {\bibfnamefont {J.}~\bibnamefont {Gu}},\ }\href@noop {} {\bibfield  {journal}
  {\bibinfo  {journal} {Chem. Phys. Chem.}\ }\textbf {\bibinfo {volume} {18}},\
  \bibinfo {pages} {101} (\bibinfo {year} {2017})}\BibitemShut {NoStop}%
\bibitem [{\citenamefont {Bagsican}\ \emph {et~al.}(2017)\citenamefont
  {Bagsican}, \citenamefont {Winchester}, \citenamefont {Ghosh}, \citenamefont
  {Zhang}, \citenamefont {Ma}, \citenamefont {Wang}, \citenamefont {Murakami},
  \citenamefont {Talapatra}, \citenamefont {Vajtai}, \citenamefont {Ajayan},
  \citenamefont {Kono}, \citenamefont {Tonouchi},\ and\ \citenamefont
  {Kawayama}}]{begsican17}%
  \BibitemOpen
  \bibfield  {author} {\bibinfo {author} {\bibfnamefont {F.~R.}\ \bibnamefont
  {Bagsican}}, \bibinfo {author} {\bibfnamefont {A.}~\bibnamefont
  {Winchester}}, \bibinfo {author} {\bibfnamefont {S.}~\bibnamefont {Ghosh}},
  \bibinfo {author} {\bibfnamefont {X.}~\bibnamefont {Zhang}}, \bibinfo
  {author} {\bibfnamefont {L.}~\bibnamefont {Ma}}, \bibinfo {author}
  {\bibfnamefont {M.}~\bibnamefont {Wang}}, \bibinfo {author} {\bibfnamefont
  {H.}~\bibnamefont {Murakami}}, \bibinfo {author} {\bibfnamefont
  {S.}~\bibnamefont {Talapatra}}, \bibinfo {author} {\bibfnamefont
  {R.}~\bibnamefont {Vajtai}}, \bibinfo {author} {\bibfnamefont {P.~M.}\
  \bibnamefont {Ajayan}}, \bibinfo {author} {\bibfnamefont {J.}~\bibnamefont
  {Kono}}, \bibinfo {author} {\bibfnamefont {M.}~\bibnamefont {Tonouchi}}, \
  and\ \bibinfo {author} {\bibfnamefont {I.}~\bibnamefont {Kawayama}},\ }\href
  {\doibase 10.1038/s41598-017-01883-1} {\bibfield  {journal} {\bibinfo
  {journal} {Sci. Rep.}\ }\textbf {\bibinfo {volume} {7}},\ \bibinfo {pages}
  {1774} (\bibinfo {year} {2017})}\BibitemShut {NoStop}%
\bibitem [{\citenamefont {Yan}\ \emph {et~al.}(2012)\citenamefont {Yan},
  \citenamefont {Xu}, \citenamefont {Shi},\ and\ \citenamefont
  {Ouyang}}]{yan12}%
  \BibitemOpen
  \bibfield  {author} {\bibinfo {author} {\bibfnamefont {H.~J.}\ \bibnamefont
  {Yan}}, \bibinfo {author} {\bibfnamefont {B.}~\bibnamefont {Xu}}, \bibinfo
  {author} {\bibfnamefont {S.~Q.}\ \bibnamefont {Shi}}, \ and\ \bibinfo
  {author} {\bibfnamefont {C.~Y.}\ \bibnamefont {Ouyang}},\ }\href@noop {}
  {\bibfield  {journal} {\bibinfo  {journal} {J.\ Appl.\ Phys.}\ }\textbf
  {\bibinfo {volume} {112}},\ \bibinfo {pages} {104316} (\bibinfo {year}
  {2012})}\BibitemShut {NoStop}%
\bibitem [{\citenamefont {Guang}\ \emph {et~al.}(2013)\citenamefont {Guang},
  \citenamefont {Aoki}, \citenamefont {Tanaka},\ and\ \citenamefont
  {Kohyama}}]{guang13}%
  \BibitemOpen
  \bibfield  {author} {\bibinfo {author} {\bibfnamefont {H.}~\bibnamefont
  {Guang}}, \bibinfo {author} {\bibfnamefont {M.}~\bibnamefont {Aoki}},
  \bibinfo {author} {\bibfnamefont {S.}~\bibnamefont {Tanaka}}, \ and\ \bibinfo
  {author} {\bibfnamefont {M.}~\bibnamefont {Kohyama}},\ }\href@noop {}
  {\bibfield  {journal} {\bibinfo  {journal} {Trans.\ Mater.\ Res.\ Soc.\
  Japan}\ }\textbf {\bibinfo {volume} {38}},\ \bibinfo {pages} {477} (\bibinfo
  {year} {2013})}\BibitemShut {NoStop}%
\bibitem [{\citenamefont {Zou}\ \emph {et~al.}(2011)\citenamefont {Zou},
  \citenamefont {Li}, \citenamefont {Zhu}, \citenamefont {Zhao}, \citenamefont
  {Xu},\ and\ \citenamefont {Su}}]{zou11}%
  \BibitemOpen
  \bibfield  {author} {\bibinfo {author} {\bibfnamefont {Y.}~\bibnamefont
  {Zou}}, \bibinfo {author} {\bibfnamefont {F.}~\bibnamefont {Li}}, \bibinfo
  {author} {\bibfnamefont {Z.~H.}\ \bibnamefont {Zhu}}, \bibinfo {author}
  {\bibfnamefont {M.~W.}\ \bibnamefont {Zhao}}, \bibinfo {author}
  {\bibfnamefont {X.~G.}\ \bibnamefont {Xu}}, \ and\ \bibinfo {author}
  {\bibfnamefont {X.~Y.}\ \bibnamefont {Su}},\ }\href@noop {} {\bibfield
  {journal} {\bibinfo  {journal} {Eur. Phys. J. B}\ }\textbf {\bibinfo {volume}
  {81}},\ \bibinfo {pages} {475} (\bibinfo {year} {2011})}\BibitemShut
  {NoStop}%
\bibitem [{\citenamefont {Kaur}\ \emph {et~al.}(2018)\citenamefont {Kaur},
  \citenamefont {Gupta}, \citenamefont {Sachdeva},\ and\ \citenamefont
  {Dharamvir}}]{kaur18}%
  \BibitemOpen
  \bibfield  {author} {\bibinfo {author} {\bibfnamefont {G.}~\bibnamefont
  {Kaur}}, \bibinfo {author} {\bibfnamefont {S.}~\bibnamefont {Gupta}},
  \bibinfo {author} {\bibfnamefont {R.}~\bibnamefont {Sachdeva}}, \ and\
  \bibinfo {author} {\bibfnamefont {K.}~\bibnamefont {Dharamvir}},\ }\href@noop
  {} {\bibfield  {journal} {\bibinfo  {journal} {AIP Conf. Proc.}\ }\textbf
  {\bibinfo {volume} {1953}},\ \bibinfo {pages} {030005} (\bibinfo {year}
  {2018})}\BibitemShut {NoStop}%
\bibitem [{\citenamefont {Benali}\ \emph {et~al.}(2014)\citenamefont {Benali},
  \citenamefont {Shulenburger}, \citenamefont {Romero}, \citenamefont {Kim},\
  and\ \citenamefont {von Lilienfeld}}]{benali14}%
  \BibitemOpen
  \bibfield  {author} {\bibinfo {author} {\bibfnamefont {A.}~\bibnamefont
  {Benali}}, \bibinfo {author} {\bibfnamefont {L.}~\bibnamefont
  {Shulenburger}}, \bibinfo {author} {\bibfnamefont {N.~A.}\ \bibnamefont
  {Romero}}, \bibinfo {author} {\bibfnamefont {J.}~\bibnamefont {Kim}}, \ and\
  \bibinfo {author} {\bibfnamefont {O.~A.}\ \bibnamefont {von Lilienfeld}},\
  }\href {\doibase 10.1021/ct5003225} {\bibfield  {journal} {\bibinfo
  {journal} {J.\ Chem.\ Theory\ Comput.}\ }\textbf {\bibinfo {volume} {10}},\
  \bibinfo {pages} {3417} (\bibinfo {year} {2014})}\BibitemShut {NoStop}%
\bibitem [{\citenamefont {Shin}\ \emph {et~al.}(2014)\citenamefont {Shin},
  \citenamefont {Kang}, \citenamefont {Koo}, \citenamefont {Lee}, \citenamefont
  {Kim},\ and\ \citenamefont {Kwon}}]{shin14}%
  \BibitemOpen
  \bibfield  {author} {\bibinfo {author} {\bibfnamefont {H.}~\bibnamefont
  {Shin}}, \bibinfo {author} {\bibfnamefont {S.}~\bibnamefont {Kang}}, \bibinfo
  {author} {\bibfnamefont {J.}~\bibnamefont {Koo}}, \bibinfo {author}
  {\bibfnamefont {H.}~\bibnamefont {Lee}}, \bibinfo {author} {\bibfnamefont
  {J.}~\bibnamefont {Kim}}, \ and\ \bibinfo {author} {\bibfnamefont
  {Y.}~\bibnamefont {Kwon}},\ }\href@noop {} {\bibfield  {journal} {\bibinfo
  {journal} {J.\ Chem.\ Phys.}\ }\textbf {\bibinfo {volume} {140}},\ \bibinfo
  {pages} {114702} (\bibinfo {year} {2014})}\BibitemShut {NoStop}%
\bibitem [{\citenamefont {Mostaani}\ \emph {et~al.}(2015)\citenamefont
  {Mostaani}, \citenamefont {Drummond},\ and\ \citenamefont
  {Fal'ko}}]{mostaani15}%
  \BibitemOpen
  \bibfield  {author} {\bibinfo {author} {\bibfnamefont {E.}~\bibnamefont
  {Mostaani}}, \bibinfo {author} {\bibfnamefont {N.~D.}\ \bibnamefont
  {Drummond}}, \ and\ \bibinfo {author} {\bibfnamefont {V.~I.}\ \bibnamefont
  {Fal'ko}},\ }\href {\doibase 10.1103/PhysRevLett.115.115501} {\bibfield
  {journal} {\bibinfo  {journal} {Phys.\ Rev.\ Lett.}\ }\textbf {\bibinfo
  {volume} {115}},\ \bibinfo {pages} {115501} (\bibinfo {year}
  {2015})}\BibitemShut {NoStop}%
\bibitem [{\citenamefont {Shin}\ \emph
  {et~al.}(2017{\natexlab{a}})\citenamefont {Shin}, \citenamefont {Kim},
  \citenamefont {Lee}, \citenamefont {Heinonen}, \citenamefont {Benali},\ and\
  \citenamefont {Kwon}}]{shin2017}%
  \BibitemOpen
  \bibfield  {author} {\bibinfo {author} {\bibfnamefont {H.}~\bibnamefont
  {Shin}}, \bibinfo {author} {\bibfnamefont {J.}~\bibnamefont {Kim}}, \bibinfo
  {author} {\bibfnamefont {H.}~\bibnamefont {Lee}}, \bibinfo {author}
  {\bibfnamefont {O.}~\bibnamefont {Heinonen}}, \bibinfo {author}
  {\bibfnamefont {A.}~\bibnamefont {Benali}}, \ and\ \bibinfo {author}
  {\bibfnamefont {Y.}~\bibnamefont {Kwon}},\ }\href {\doibase
  10.1021/acs.jctc.7b00747} {\bibfield  {journal} {\bibinfo  {journal} {J.
  Chem. Theory Comput.}\ }\textbf {\bibinfo {volume} {13}},\ \bibinfo {pages}
  {5639} (\bibinfo {year} {2017}{\natexlab{a}})}\BibitemShut {NoStop}%
\bibitem [{\citenamefont {Reynolds}\ \emph {et~al.}(1982)\citenamefont
  {Reynolds}, \citenamefont {Ceperley}, \citenamefont {Alder},\ and\
  \citenamefont {Lester}}]{Reynolds1982}%
  \BibitemOpen
  \bibfield  {author} {\bibinfo {author} {\bibfnamefont {P.~J.}\ \bibnamefont
  {Reynolds}}, \bibinfo {author} {\bibfnamefont {D.~M.}\ \bibnamefont
  {Ceperley}}, \bibinfo {author} {\bibfnamefont {B.~J.}\ \bibnamefont {Alder}},
  \ and\ \bibinfo {author} {\bibfnamefont {W.~A.}\ \bibnamefont {Lester}},\
  }\href {\doibase 10.1063/1.443766} {\bibfield  {journal} {\bibinfo  {journal}
  {J. Chem. Phys.}\ }\textbf {\bibinfo {volume} {77}},\ \bibinfo {pages} {5593}
  (\bibinfo {year} {1982})}\BibitemShut {NoStop}%
\bibitem [{\citenamefont {Foulkes}\ \emph {et~al.}(2001)\citenamefont
  {Foulkes}, \citenamefont {Mitas}, \citenamefont {Needs},\ and\ \citenamefont
  {Rajagopal}}]{foulkes01}%
  \BibitemOpen
  \bibfield  {author} {\bibinfo {author} {\bibfnamefont {W.~M.~C.}\
  \bibnamefont {Foulkes}}, \bibinfo {author} {\bibfnamefont {L.}~\bibnamefont
  {Mitas}}, \bibinfo {author} {\bibfnamefont {R.~J.}\ \bibnamefont {Needs}}, \
  and\ \bibinfo {author} {\bibfnamefont {G.}~\bibnamefont {Rajagopal}},\
  }\href@noop {} {\bibfield  {journal} {\bibinfo  {journal} {Rev.\ Mod.\
  Phys.}\ }\textbf {\bibinfo {volume} {73}},\ \bibinfo {pages} {33} (\bibinfo
  {year} {2001})}\BibitemShut {NoStop}%
\bibitem [{\citenamefont {Kim}\ \emph {et~al.}(2018)\citenamefont {Kim},
  \citenamefont {Baczewski}, \citenamefont {Beaudet}, \citenamefont {Benali},
  \citenamefont {Bennett}, \citenamefont {Berrill}, \citenamefont {Blunt},
  \citenamefont {Borda}, \citenamefont {Casula}, \citenamefont {Ceperley},
  \citenamefont {Chiesa}, \citenamefont {Clark}, \citenamefont {Clay},
  \citenamefont {Delaney}, \citenamefont {Dewing}, \citenamefont {Esler},
  \citenamefont {Hao}, \citenamefont {Heinonen}, \citenamefont {Kent},
  \citenamefont {Krogel}, \citenamefont {Kylanpaa}, \citenamefont {Li},
  \citenamefont {Lopez}, \citenamefont {Luo}, \citenamefont {Malone},
  \citenamefont {Martin}, \citenamefont {Mathuriya}, \citenamefont {McMinis},
  \citenamefont {Melton}, \citenamefont {Mitas}, \citenamefont {Morales},
  \citenamefont {Neuscamman}, \citenamefont {Parker}, \citenamefont {Flores},
  \citenamefont {Romero}, \citenamefont {Rubenstein}, \citenamefont {Shea},
  \citenamefont {Shin}, \citenamefont {Shulenburger}, \citenamefont {Tillack},
  \citenamefont {Townsend}, \citenamefont {Tubman}, \citenamefont {van~der
  Goetz}, \citenamefont {Vincent}, \citenamefont {Yang}, \citenamefont {Yang},
  \citenamefont {Zhang},\ and\ \citenamefont {Zhao}}]{QMCPACK}%
  \BibitemOpen
  \bibfield  {author} {\bibinfo {author} {\bibfnamefont {J.}~\bibnamefont
  {Kim}}, \bibinfo {author} {\bibfnamefont {A.}~\bibnamefont {Baczewski}},
  \bibinfo {author} {\bibfnamefont {T.}~\bibnamefont {Beaudet}}, \bibinfo
  {author} {\bibfnamefont {A.}~\bibnamefont {Benali}}, \bibinfo {author}
  {\bibfnamefont {C.}~\bibnamefont {Bennett}}, \bibinfo {author} {\bibfnamefont
  {M.}~\bibnamefont {Berrill}}, \bibinfo {author} {\bibfnamefont
  {N.}~\bibnamefont {Blunt}}, \bibinfo {author} {\bibfnamefont {E.~J.~L.}\
  \bibnamefont {Borda}}, \bibinfo {author} {\bibfnamefont {M.}~\bibnamefont
  {Casula}}, \bibinfo {author} {\bibfnamefont {D.}~\bibnamefont {Ceperley}},
  \bibinfo {author} {\bibfnamefont {S.}~\bibnamefont {Chiesa}}, \bibinfo
  {author} {\bibfnamefont {B.~K.}\ \bibnamefont {Clark}}, \bibinfo {author}
  {\bibfnamefont {R.}~\bibnamefont {Clay}}, \bibinfo {author} {\bibfnamefont
  {K.}~\bibnamefont {Delaney}}, \bibinfo {author} {\bibfnamefont
  {M.}~\bibnamefont {Dewing}}, \bibinfo {author} {\bibfnamefont
  {K.}~\bibnamefont {Esler}}, \bibinfo {author} {\bibfnamefont
  {H.}~\bibnamefont {Hao}}, \bibinfo {author} {\bibfnamefont {O.}~\bibnamefont
  {Heinonen}}, \bibinfo {author} {\bibfnamefont {P.~R.~C.}\ \bibnamefont
  {Kent}}, \bibinfo {author} {\bibfnamefont {J.~T.}\ \bibnamefont {Krogel}},
  \bibinfo {author} {\bibfnamefont {I.}~\bibnamefont {Kylanpaa}}, \bibinfo
  {author} {\bibfnamefont {Y.~W.}\ \bibnamefont {Li}}, \bibinfo {author}
  {\bibfnamefont {M.~G.}\ \bibnamefont {Lopez}}, \bibinfo {author}
  {\bibfnamefont {Y.}~\bibnamefont {Luo}}, \bibinfo {author} {\bibfnamefont
  {F.}~\bibnamefont {Malone}}, \bibinfo {author} {\bibfnamefont
  {R.}~\bibnamefont {Martin}}, \bibinfo {author} {\bibfnamefont
  {A.}~\bibnamefont {Mathuriya}}, \bibinfo {author} {\bibfnamefont
  {J.}~\bibnamefont {McMinis}}, \bibinfo {author} {\bibfnamefont
  {C.}~\bibnamefont {Melton}}, \bibinfo {author} {\bibfnamefont
  {L.}~\bibnamefont {Mitas}}, \bibinfo {author} {\bibfnamefont {M.~A.}\
  \bibnamefont {Morales}}, \bibinfo {author} {\bibfnamefont {E.}~\bibnamefont
  {Neuscamman}}, \bibinfo {author} {\bibfnamefont {W.}~\bibnamefont {Parker}},
  \bibinfo {author} {\bibfnamefont {S.}~\bibnamefont {Flores}}, \bibinfo
  {author} {\bibfnamefont {N.~A.}\ \bibnamefont {Romero}}, \bibinfo {author}
  {\bibfnamefont {B.}~\bibnamefont {Rubenstein}}, \bibinfo {author}
  {\bibfnamefont {J.}~\bibnamefont {Shea}}, \bibinfo {author} {\bibfnamefont
  {H.}~\bibnamefont {Shin}}, \bibinfo {author} {\bibfnamefont {L.}~\bibnamefont
  {Shulenburger}}, \bibinfo {author} {\bibfnamefont {A.}~\bibnamefont
  {Tillack}}, \bibinfo {author} {\bibfnamefont {J.}~\bibnamefont {Townsend}},
  \bibinfo {author} {\bibfnamefont {N.}~\bibnamefont {Tubman}}, \bibinfo
  {author} {\bibfnamefont {B.}~\bibnamefont {van~der Goetz}}, \bibinfo {author}
  {\bibfnamefont {J.}~\bibnamefont {Vincent}}, \bibinfo {author} {\bibfnamefont
  {D.~C.}\ \bibnamefont {Yang}}, \bibinfo {author} {\bibfnamefont
  {Y.}~\bibnamefont {Yang}}, \bibinfo {author} {\bibfnamefont {S.}~\bibnamefont
  {Zhang}}, \ and\ \bibinfo {author} {\bibfnamefont {L.}~\bibnamefont {Zhao}},\
  }\href {\doibase 10.1088/1361-648X/aab9c3} {\bibfield  {journal} {\bibinfo
  {journal} {J. Phys.: Condens. Matter}\ }\textbf {\bibinfo {volume} {30}},\
  \bibinfo {pages} {195901} (\bibinfo {year} {2018})}\BibitemShut {NoStop}%
\bibitem [{\citenamefont {Casula}\ \emph {et~al.}(2010)\citenamefont {Casula},
  \citenamefont {Moroni}, \citenamefont {Sorella},\ and\ \citenamefont
  {Filippi}}]{casula10}%
  \BibitemOpen
  \bibfield  {author} {\bibinfo {author} {\bibfnamefont {M.}~\bibnamefont
  {Casula}}, \bibinfo {author} {\bibfnamefont {S.}~\bibnamefont {Moroni}},
  \bibinfo {author} {\bibfnamefont {S.}~\bibnamefont {Sorella}}, \ and\
  \bibinfo {author} {\bibfnamefont {C.}~\bibnamefont {Filippi}},\ }\href
  {\doibase 10.1063/1.3380831} {\bibfield  {journal} {\bibinfo  {journal} {J.\
  Chem.\ Phys.}\ }\textbf {\bibinfo {volume} {132}},\ \bibinfo {pages} {154113}
  (\bibinfo {year} {2010})}\BibitemShut {NoStop}%
\bibitem [{\citenamefont {Lin}\ \emph {et~al.}(2001)\citenamefont {Lin},
  \citenamefont {Zong},\ and\ \citenamefont {Ceperley}}]{lin01}%
  \BibitemOpen
  \bibfield  {author} {\bibinfo {author} {\bibfnamefont {C.}~\bibnamefont
  {Lin}}, \bibinfo {author} {\bibfnamefont {F.~H.}\ \bibnamefont {Zong}}, \
  and\ \bibinfo {author} {\bibfnamefont {D.~M.}\ \bibnamefont {Ceperley}},\
  }\href@noop {} {\bibfield  {journal} {\bibinfo  {journal} {Phys.\ Rev.\ E}\
  }\textbf {\bibinfo {volume} {64}},\ \bibinfo {pages} {016702} (\bibinfo
  {year} {2001})}\BibitemShut {NoStop}%
\bibitem [{\citenamefont {Burkatzki}\ \emph {et~al.}(2007)\citenamefont
  {Burkatzki}, \citenamefont {Filippi},\ and\ \citenamefont
  {Dolg}}]{burkatzki07}%
  \BibitemOpen
  \bibfield  {author} {\bibinfo {author} {\bibfnamefont {M.}~\bibnamefont
  {Burkatzki}}, \bibinfo {author} {\bibfnamefont {C.}~\bibnamefont {Filippi}},
  \ and\ \bibinfo {author} {\bibfnamefont {M.}~\bibnamefont {Dolg}},\
  }\href@noop {} {\bibfield  {journal} {\bibinfo  {journal} {J.\ Chem.\ Phys.}\
  }\textbf {\bibinfo {volume} {126}},\ \bibinfo {pages} {234105} (\bibinfo
  {year} {2007})}\BibitemShut {NoStop}%
\bibitem [{\citenamefont {Burkatzki}\ \emph {et~al.}(2008)\citenamefont
  {Burkatzki}, \citenamefont {Filippi},\ and\ \citenamefont
  {Dolg}}]{burkatzki08}%
  \BibitemOpen
  \bibfield  {author} {\bibinfo {author} {\bibfnamefont {M.}~\bibnamefont
  {Burkatzki}}, \bibinfo {author} {\bibfnamefont {C.}~\bibnamefont {Filippi}},
  \ and\ \bibinfo {author} {\bibfnamefont {M.}~\bibnamefont {Dolg}},\
  }\href@noop {} {\bibfield  {journal} {\bibinfo  {journal} {J.\ Chem.\ Phys.}\
  }\textbf {\bibinfo {volume} {129}},\ \bibinfo {pages} {164115} (\bibinfo
  {year} {2008})}\BibitemShut {NoStop}%
\bibitem [{\citenamefont {Benali}\ \emph {et~al.}(2016)\citenamefont {Benali},
  \citenamefont {Shulenburger}, \citenamefont {Krogel}, \citenamefont {Zhong},
  \citenamefont {Kent},\ and\ \citenamefont {Heinonen}}]{benali16}%
  \BibitemOpen
  \bibfield  {author} {\bibinfo {author} {\bibfnamefont {A.}~\bibnamefont
  {Benali}}, \bibinfo {author} {\bibfnamefont {L.}~\bibnamefont
  {Shulenburger}}, \bibinfo {author} {\bibfnamefont {J.~T.}\ \bibnamefont
  {Krogel}}, \bibinfo {author} {\bibfnamefont {X.}~\bibnamefont {Zhong}},
  \bibinfo {author} {\bibfnamefont {P.~R.~C.}\ \bibnamefont {Kent}}, \ and\
  \bibinfo {author} {\bibfnamefont {O.}~\bibnamefont {Heinonen}},\ }\href@noop
  {} {\bibfield  {journal} {\bibinfo  {journal} {Phys.\ Chem.\ Chem.\ Phys.}\
  }\textbf {\bibinfo {volume} {18}},\ \bibinfo {pages} {18323} (\bibinfo {year}
  {2016})}\BibitemShut {NoStop}%
\bibitem [{\citenamefont {Luo}\ \emph {et~al.}(2016)\citenamefont {Luo},
  \citenamefont {Benali}, \citenamefont {Shulenburger}, \citenamefont {Krogel},
  \citenamefont {Heinonen},\ and\ \citenamefont {Kent}}]{luo16}%
  \BibitemOpen
  \bibfield  {author} {\bibinfo {author} {\bibfnamefont {Y.}~\bibnamefont
  {Luo}}, \bibinfo {author} {\bibfnamefont {A.}~\bibnamefont {Benali}},
  \bibinfo {author} {\bibfnamefont {L.}~\bibnamefont {Shulenburger}}, \bibinfo
  {author} {\bibfnamefont {J.~T.}\ \bibnamefont {Krogel}}, \bibinfo {author}
  {\bibfnamefont {O.}~\bibnamefont {Heinonen}}, \ and\ \bibinfo {author}
  {\bibfnamefont {P.~R.~C.}\ \bibnamefont {Kent}},\ }\href@noop {} {\bibfield
  {journal} {\bibinfo  {journal} {New\ J.\ Phys.}\ }\textbf {\bibinfo {volume}
  {18}},\ \bibinfo {pages} {113049} (\bibinfo {year} {2016})}\BibitemShut
  {NoStop}%
\bibitem [{\citenamefont {Shin}\ \emph
  {et~al.}(2017{\natexlab{b}})\citenamefont {Shin}, \citenamefont {Luo},
  \citenamefont {Ganesh}, \citenamefont {Balachandran}, \citenamefont {Krogel},
  \citenamefont {Kent}, \citenamefont {Benali},\ and\ \citenamefont
  {Heinonen}}]{shin17}%
  \BibitemOpen
  \bibfield  {author} {\bibinfo {author} {\bibfnamefont {H.}~\bibnamefont
  {Shin}}, \bibinfo {author} {\bibfnamefont {Y.}~\bibnamefont {Luo}}, \bibinfo
  {author} {\bibfnamefont {P.}~\bibnamefont {Ganesh}}, \bibinfo {author}
  {\bibfnamefont {J.}~\bibnamefont {Balachandran}}, \bibinfo {author}
  {\bibfnamefont {J.~T.}\ \bibnamefont {Krogel}}, \bibinfo {author}
  {\bibfnamefont {P.~R.~C.}\ \bibnamefont {Kent}}, \bibinfo {author}
  {\bibfnamefont {A.}~\bibnamefont {Benali}}, \ and\ \bibinfo {author}
  {\bibfnamefont {O.}~\bibnamefont {Heinonen}},\ }\href@noop {} {\bibfield
  {journal} {\bibinfo  {journal} {Phys.\ Rev.\ Materials}\ }\textbf {\bibinfo
  {volume} {1}},\ \bibinfo {pages} {173603} (\bibinfo {year}
  {2017}{\natexlab{b}})}\BibitemShut {NoStop}%
\bibitem [{\citenamefont {Perdew}\ \emph {et~al.}(1996)\citenamefont {Perdew},
  \citenamefont {Burke},\ and\ \citenamefont {Ernzerhof}}]{perdew96}%
  \BibitemOpen
  \bibfield  {author} {\bibinfo {author} {\bibfnamefont {J.~P.}\ \bibnamefont
  {Perdew}}, \bibinfo {author} {\bibfnamefont {K.}~\bibnamefont {Burke}}, \
  and\ \bibinfo {author} {\bibfnamefont {M.}~\bibnamefont {Ernzerhof}},\
  }\href@noop {} {\bibfield  {journal} {\bibinfo  {journal} {Phys.\ Rev.\
  Lett.}\ }\textbf {\bibinfo {volume} {77}},\ \bibinfo {pages} {3865} (\bibinfo
  {year} {1996})}\BibitemShut {NoStop}%
\bibitem [{\citenamefont {Giannozzi}\ \emph {et~al.}(2009)\citenamefont
  {Giannozzi}, \citenamefont {Baroni}, \citenamefont {Bonini}, \citenamefont
  {Calandra}, \citenamefont {Car}, \citenamefont {Cavazzoni}, \citenamefont
  {Ceresoli}, \citenamefont {Chiarotti}, \citenamefont {Cococcioni},
  \citenamefont {Dabo}, \citenamefont {Corso}, \citenamefont {de~Gironcoli},
  \citenamefont {Fabris}, \citenamefont {Fratesi}, \citenamefont {Gebauer},
  \citenamefont {Gerstmann}, \citenamefont {Gougoussis}, \citenamefont
  {Kokalj}, \citenamefont {Lazzeri}, \citenamefont {M-.Samos}, \citenamefont
  {Mazari}, \citenamefont {Mauri}, \citenamefont {Mazzarello}, \citenamefont
  {Paolini}, \citenamefont {Pasquarello}, \citenamefont {Paulatto},
  \citenamefont {Sbraccia}, \citenamefont {Scandolo}, \citenamefont
  {Sclauzero}, \citenamefont {Seitsonen}, \citenamefont {Smogunov},
  \citenamefont {Umari},\ and\ \citenamefont {Wentzcovitch}}]{QE}%
  \BibitemOpen
  \bibfield  {author} {\bibinfo {author} {\bibfnamefont {P.}~\bibnamefont
  {Giannozzi}}, \bibinfo {author} {\bibfnamefont {S.}~\bibnamefont {Baroni}},
  \bibinfo {author} {\bibfnamefont {N.}~\bibnamefont {Bonini}}, \bibinfo
  {author} {\bibfnamefont {M.}~\bibnamefont {Calandra}}, \bibinfo {author}
  {\bibfnamefont {R.}~\bibnamefont {Car}}, \bibinfo {author} {\bibfnamefont
  {C.}~\bibnamefont {Cavazzoni}}, \bibinfo {author} {\bibfnamefont
  {D.}~\bibnamefont {Ceresoli}}, \bibinfo {author} {\bibfnamefont {G.~L.}\
  \bibnamefont {Chiarotti}}, \bibinfo {author} {\bibfnamefont {M.}~\bibnamefont
  {Cococcioni}}, \bibinfo {author} {\bibfnamefont {I.}~\bibnamefont {Dabo}},
  \bibinfo {author} {\bibfnamefont {A.~D.}\ \bibnamefont {Corso}}, \bibinfo
  {author} {\bibfnamefont {S.}~\bibnamefont {de~Gironcoli}}, \bibinfo {author}
  {\bibfnamefont {S.}~\bibnamefont {Fabris}}, \bibinfo {author} {\bibfnamefont
  {G.}~\bibnamefont {Fratesi}}, \bibinfo {author} {\bibfnamefont
  {R.}~\bibnamefont {Gebauer}}, \bibinfo {author} {\bibfnamefont
  {U.}~\bibnamefont {Gerstmann}}, \bibinfo {author} {\bibfnamefont
  {C.}~\bibnamefont {Gougoussis}}, \bibinfo {author} {\bibfnamefont
  {A.}~\bibnamefont {Kokalj}}, \bibinfo {author} {\bibfnamefont
  {M.}~\bibnamefont {Lazzeri}}, \bibinfo {author} {\bibfnamefont
  {L.}~\bibnamefont {M-.Samos}}, \bibinfo {author} {\bibfnamefont
  {N.}~\bibnamefont {Mazari}}, \bibinfo {author} {\bibfnamefont
  {F.}~\bibnamefont {Mauri}}, \bibinfo {author} {\bibfnamefont
  {R.}~\bibnamefont {Mazzarello}}, \bibinfo {author} {\bibfnamefont
  {S.}~\bibnamefont {Paolini}}, \bibinfo {author} {\bibfnamefont
  {A.}~\bibnamefont {Pasquarello}}, \bibinfo {author} {\bibfnamefont
  {L.}~\bibnamefont {Paulatto}}, \bibinfo {author} {\bibfnamefont
  {C.}~\bibnamefont {Sbraccia}}, \bibinfo {author} {\bibfnamefont
  {S.}~\bibnamefont {Scandolo}}, \bibinfo {author} {\bibfnamefont
  {G.}~\bibnamefont {Sclauzero}}, \bibinfo {author} {\bibfnamefont {A.~P.}\
  \bibnamefont {Seitsonen}}, \bibinfo {author} {\bibfnamefont {A.}~\bibnamefont
  {Smogunov}}, \bibinfo {author} {\bibfnamefont {P.}~\bibnamefont {Umari}}, \
  and\ \bibinfo {author} {\bibfnamefont {R.~M.}\ \bibnamefont {Wentzcovitch}},\
  }\href@noop {} {\bibfield  {journal} {\bibinfo  {journal} {J.\ Phys.:\
  Condens.\ Matter}\ }\textbf {\bibinfo {volume} {21}},\ \bibinfo {pages}
  {395502} (\bibinfo {year} {2009})}\BibitemShut {NoStop}%
\bibitem [{\citenamefont {Grimme}(2004)}]{grimme04}%
  \BibitemOpen
  \bibfield  {author} {\bibinfo {author} {\bibfnamefont {S.}~\bibnamefont
  {Grimme}},\ }\href {\doibase 10.1002/jcc.20078} {\bibfield  {journal}
  {\bibinfo  {journal} {J.\ Comput.\ Chem.}\ }\textbf {\bibinfo {volume}
  {25}},\ \bibinfo {pages} {1463} (\bibinfo {year} {2004})}\BibitemShut
  {NoStop}%
\bibitem [{\citenamefont {Grimme}(2006)}]{grimme06}%
  \BibitemOpen
  \bibfield  {author} {\bibinfo {author} {\bibfnamefont {S.}~\bibnamefont
  {Grimme}},\ }\href {\doibase 10.1002/jcc.20495} {\bibfield  {journal}
  {\bibinfo  {journal} {J.\ Comput.\ Chem.}\ }\textbf {\bibinfo {volume}
  {27}},\ \bibinfo {pages} {1787} (\bibinfo {year} {2006})}\BibitemShut
  {NoStop}%
\bibitem [{\citenamefont {Grimme}\ \emph {et~al.}(2010)\citenamefont {Grimme},
  \citenamefont {Antony}, \citenamefont {Ehrlich},\ and\ \citenamefont
  {Krieg}}]{grimme10}%
  \BibitemOpen
  \bibfield  {author} {\bibinfo {author} {\bibfnamefont {S.}~\bibnamefont
  {Grimme}}, \bibinfo {author} {\bibfnamefont {J.}~\bibnamefont {Antony}},
  \bibinfo {author} {\bibfnamefont {S.}~\bibnamefont {Ehrlich}}, \ and\
  \bibinfo {author} {\bibfnamefont {H.}~\bibnamefont {Krieg}},\ }\href
  {\doibase 10.1063/1.3382344} {\bibfield  {journal} {\bibinfo  {journal} {J.\
  Chem.\ Phys.}\ }\textbf {\bibinfo {volume} {132}},\ \bibinfo {pages} {154104}
  (\bibinfo {year} {2010})}\BibitemShut {NoStop}%
\bibitem [{\citenamefont {Dion}\ \emph {et~al.}(2004)\citenamefont {Dion},
  \citenamefont {Rydberg}, \citenamefont {Schr$\text{\"{o}}$der}, \citenamefont
  {Langreth},\ and\ \citenamefont {Lundqvist}}]{dion04}%
  \BibitemOpen
  \bibfield  {author} {\bibinfo {author} {\bibfnamefont {M.}~\bibnamefont
  {Dion}}, \bibinfo {author} {\bibfnamefont {H.}~\bibnamefont {Rydberg}},
  \bibinfo {author} {\bibfnamefont {E.}~\bibnamefont {Schr$\text{\"{o}}$der}},
  \bibinfo {author} {\bibfnamefont {D.~C.}\ \bibnamefont {Langreth}}, \ and\
  \bibinfo {author} {\bibfnamefont {B.~I.}\ \bibnamefont {Lundqvist}},\ }\href
  {\doibase 10.1103/PhysRevLett.92.246401} {\bibfield  {journal} {\bibinfo
  {journal} {Phys.\ Rev.\ Lett.}\ }\textbf {\bibinfo {volume} {92}},\ \bibinfo
  {pages} {246401} (\bibinfo {year} {2004})}\BibitemShut {NoStop}%
\bibitem [{\citenamefont {Lee}\ \emph {et~al.}(2010)\citenamefont {Lee},
  \citenamefont {$\text{\'{E}}$. D.~Murray}, \citenamefont {Kong},
  \citenamefont {Lundqvist},\ and\ \citenamefont {Langreth}}]{lee10}%
  \BibitemOpen
  \bibfield  {author} {\bibinfo {author} {\bibfnamefont {K.}~\bibnamefont
  {Lee}}, \bibinfo {author} {\bibnamefont {$\text{\'{E}}$. D.~Murray}},
  \bibinfo {author} {\bibfnamefont {L.}~\bibnamefont {Kong}}, \bibinfo {author}
  {\bibfnamefont {B.~I.}\ \bibnamefont {Lundqvist}}, \ and\ \bibinfo {author}
  {\bibfnamefont {D.~C.}\ \bibnamefont {Langreth}},\ }\href {\doibase
  10.1103/PhysRevB.82.081101} {\bibfield  {journal} {\bibinfo  {journal}
  {Phys.\ Rev.\ B}\ }\textbf {\bibinfo {volume} {82}},\ \bibinfo {pages}
  {081101(R)} (\bibinfo {year} {2010})}\BibitemShut {NoStop}%
\bibitem [{\citenamefont {Sabatini}\ \emph {et~al.}(2013)\citenamefont
  {Sabatini}, \citenamefont {Gorni},\ and\ \citenamefont
  {de~Gironcoli}}]{sabatini13}%
  \BibitemOpen
  \bibfield  {author} {\bibinfo {author} {\bibfnamefont {R.}~\bibnamefont
  {Sabatini}}, \bibinfo {author} {\bibfnamefont {T.}~\bibnamefont {Gorni}}, \
  and\ \bibinfo {author} {\bibfnamefont {S.}~\bibnamefont {de~Gironcoli}},\
  }\href {\doibase 10.1103/PhysRevB.87.041108} {\bibfield  {journal} {\bibinfo
  {journal} {Phys.\ Rev.\ B}\ }\textbf {\bibinfo {volume} {87}},\ \bibinfo
  {pages} {041108(R)} (\bibinfo {year} {2013})}\BibitemShut {NoStop}%
\bibitem [{\citenamefont {Aquilanti}\ \emph {et~al.}(1999)\citenamefont
  {Aquilanti}, \citenamefont {Ascenzi}, \citenamefont {Bartolomei},
  \citenamefont {Cappelletti}, \citenamefont {Cavalli}, \citenamefont
  {de~Castro~Vitores},\ and\ \citenamefont {Pirani}}]{aquilanti99}%
  \BibitemOpen
  \bibfield  {author} {\bibinfo {author} {\bibfnamefont {V.}~\bibnamefont
  {Aquilanti}}, \bibinfo {author} {\bibfnamefont {D.}~\bibnamefont {Ascenzi}},
  \bibinfo {author} {\bibfnamefont {M.}~\bibnamefont {Bartolomei}}, \bibinfo
  {author} {\bibfnamefont {D.}~\bibnamefont {Cappelletti}}, \bibinfo {author}
  {\bibfnamefont {S.}~\bibnamefont {Cavalli}}, \bibinfo {author} {\bibfnamefont
  {M.}~\bibnamefont {de~Castro~Vitores}}, \ and\ \bibinfo {author}
  {\bibfnamefont {F.}~\bibnamefont {Pirani}},\ }\href {\doibase
  10.1021/ja9917215} {\bibfield  {journal} {\bibinfo  {journal} {J. Am. Chem.
  Soc.}\ }\textbf {\bibinfo {volume} {121}},\ \bibinfo {pages} {10794}
  (\bibinfo {year} {1999})}\BibitemShut {NoStop}%
\bibitem [{\citenamefont {Hern\'{a}ndez-Lamoneda}\ \emph
  {et~al.}(2005)\citenamefont {Hern\'{a}ndez-Lamoneda}, \citenamefont
  {Bartolomei}, \citenamefont {Hern\'{a}ndez}, \citenamefont
  {Campos-Martinez},\ and\ \citenamefont {Dayou}}]{lamoneda05}%
  \BibitemOpen
  \bibfield  {author} {\bibinfo {author} {\bibfnamefont {R.}~\bibnamefont
  {Hern\'{a}ndez-Lamoneda}}, \bibinfo {author} {\bibfnamefont {M.}~\bibnamefont
  {Bartolomei}}, \bibinfo {author} {\bibfnamefont {M.~I.}\ \bibnamefont
  {Hern\'{a}ndez}}, \bibinfo {author} {\bibfnamefont {J.}~\bibnamefont
  {Campos-Martinez}}, \ and\ \bibinfo {author} {\bibfnamefont {F.}~\bibnamefont
  {Dayou}},\ }\href {\doibase 10.1021/jp053728g} {\bibfield  {journal}
  {\bibinfo  {journal} {J. Phys. Chem. A}\ }\textbf {\bibinfo {volume} {109}},\
  \bibinfo {pages} {11587} (\bibinfo {year} {2005})}\BibitemShut {NoStop}%
\bibitem [{\citenamefont {Monkhorst}\ and\ \citenamefont
  {Pack}(1976)}]{monkhorst76}%
  \BibitemOpen
  \bibfield  {author} {\bibinfo {author} {\bibfnamefont {H.~J.}\ \bibnamefont
  {Monkhorst}}\ and\ \bibinfo {author} {\bibfnamefont {J.~D.}\ \bibnamefont
  {Pack}},\ }\href@noop {} {\bibfield  {journal} {\bibinfo  {journal} {Phys.\
  Rev.\ B}\ }\textbf {\bibinfo {volume} {13}},\ \bibinfo {pages} {5188}
  (\bibinfo {year} {1976})}\BibitemShut {NoStop}%
\bibitem [{\citenamefont {Lide}(2003)}]{lide03}%
  \BibitemOpen
  \bibfield  {author} {\bibinfo {author} {\bibfnamefont {D.~R.}\ \bibnamefont
  {Lide}},\ }\href@noop {} {\emph {\bibinfo {title} {CRC Handbook of Chemistry
  and Physics, 84th Ed.}}}\ (\bibinfo  {publisher} {CRC Press},\ \bibinfo
  {year} {2003})\BibitemShut {NoStop}%
\bibitem [{\citenamefont {Lynch}\ and\ \citenamefont
  {Drickamer}(1966)}]{lynch66}%
  \BibitemOpen
  \bibfield  {author} {\bibinfo {author} {\bibfnamefont {R.~W.}\ \bibnamefont
  {Lynch}}\ and\ \bibinfo {author} {\bibfnamefont {H.~G.}\ \bibnamefont
  {Drickamer}},\ }\href@noop {} {\bibfield  {journal} {\bibinfo  {journal} {J.
  Chem. Phys.}\ }\textbf {\bibinfo {volume} {44}},\ \bibinfo {pages} {181}
  (\bibinfo {year} {1966})}\BibitemShut {NoStop}%
\bibitem [{Note1()}]{Note1}%
  \BibitemOpen
  \bibinfo {note} {See Supplemental Material at [URL will be inserted by
  publisher] for more information on finite-size analysis for the adsorption
  energy of O$_2$ on graphene.}\BibitemShut {Stop}%
\bibitem [{\citenamefont {Shulenburger}\ \emph {et~al.}(2015)\citenamefont
  {Shulenburger}, \citenamefont {Baczewski}, \citenamefont {Zhu}, \citenamefont
  {Guan},\ and\ \citenamefont {Tom$\text{\'{a}}$nek}}]{shulenburger15}%
  \BibitemOpen
  \bibfield  {author} {\bibinfo {author} {\bibfnamefont {L.}~\bibnamefont
  {Shulenburger}}, \bibinfo {author} {\bibfnamefont {A.~D.}\ \bibnamefont
  {Baczewski}}, \bibinfo {author} {\bibfnamefont {Z.}~\bibnamefont {Zhu}},
  \bibinfo {author} {\bibfnamefont {J.}~\bibnamefont {Guan}}, \ and\ \bibinfo
  {author} {\bibfnamefont {D.}~\bibnamefont {Tom$\text{\'{a}}$nek}},\ }\href
  {\doibase 10.1021/acs.nanolett.5b03615} {\bibfield  {journal} {\bibinfo
  {journal} {Nano\ Lett.}\ }\textbf {\bibinfo {volume} {15}},\ \bibinfo {pages}
  {8170} (\bibinfo {year} {2015})}\BibitemShut {NoStop}%
\bibitem [{\citenamefont {Ahn}\ \emph {et~al.}(2018)\citenamefont {Ahn},
  \citenamefont {Hong}, \citenamefont {Kwon}, \citenamefont {Clay},
  \citenamefont {Shulenburger}, \citenamefont {Shin},\ and\ \citenamefont
  {Benali}}]{ahn18}%
  \BibitemOpen
  \bibfield  {author} {\bibinfo {author} {\bibfnamefont {J.}~\bibnamefont
  {Ahn}}, \bibinfo {author} {\bibfnamefont {I.}~\bibnamefont {Hong}}, \bibinfo
  {author} {\bibfnamefont {Y.}~\bibnamefont {Kwon}}, \bibinfo {author}
  {\bibfnamefont {R.~C.}\ \bibnamefont {Clay}}, \bibinfo {author}
  {\bibfnamefont {L.}~\bibnamefont {Shulenburger}}, \bibinfo {author}
  {\bibfnamefont {H.}~\bibnamefont {Shin}}, \ and\ \bibinfo {author}
  {\bibfnamefont {A.}~\bibnamefont {Benali}},\ }\href@noop {} {\bibfield
  {journal} {\bibinfo  {journal} {Phys. Rev. B}\ }\textbf {\bibinfo {volume}
  {98}},\ \bibinfo {pages} {085429} (\bibinfo {year} {2018})}\BibitemShut
  {NoStop}%
\end{thebibliography}%

\end{document}